# Mechanical properties and deformation mechanisms of the C14 Laves and μ-phase in the ternary Ta-Fe(-Al) system

C. Gasper[1]*, E.M. Soysal[1], N. Ulumuddin[1], T. Stollenwerk[1], T. Reclik[1], P.L. Sun[1], S. Korte-Kerzel[1]

[1]Institute for Physical Metallurgy and Materials Physics, RWTH Aachen University, 52074 Aachen, Germany

*gasper@imm.rwth-aachen.de (corresponding author)

## Abstract

As structural and functional materials, topologically close-packed (TCP) phases of transition metal compounds offer a wide range of attractive properties. Due to their complex crystal structure and the resulting brittleness, the knowledge on their mechanical behaviour is still very limited, especially below the brittle-to-ductile transition temperature. In this study, we systematically analyse the influence of composition and crystal structure on the mechanical properties and deformation mechanisms in the binary Ta-Fe system as well as in the ternary Ta-Fe-Al system as both systems contain a hexagonal C14 Laves and a μ-phase. We use nanoindentation, slip trace analysis and transmission electron microscopy to study the influence of crystal structure, composition and crystal orientation. The composition strongly influences the indentation modulus in the binary Ta-Fe system, showing a decreasing trend with increasing Ta content. The addition of Al, however, does not lead to a significant change of the mechanical properties of the ternary TCP phases. The investigation of the deformation mechanisms revealed that the Laves phase primarily deforms via non-basal slip while the basal plane is the favoured slip plane in the μ-phase. By partly replacing Fe with Al, the plasticity is not affected strongly, but the proportion of non-basal slip slightly increases for both ternary TCP phases compared to the binary ones.

# 1 Introduction

First studies on the crystal structure of intermetallic compounds, consisting of at least two different elements and with a stoichiometric composition of $AB_2$ (with A denoting the larger atoms and B the smaller ones), such as $MgZn_2$ and $MgCu_2$, were performed by Friauf in 1927 [1, 2]. In the 1930s, Laves et al. [3–6] further investigated the crystallographic similarity of different intermetallic compounds that can be classified as the cubic $MgCu_2$ and the hexagonal $MgZn_2$ and $MgNi_2$ structure types and have an ideal radius ratio of the component atoms of $\sqrt{3/2} \approx 1.225$. The term "Laves phases" was first used by Schulze [7], who also studied the structure of Laves phases, the relationships between the different structure types and their differentiation compared to other intermetallic phases. Frank and Kasper [8, 9] found later in the 1950s that also these complex close-packed structures can be grouped into four polyhedra with triangular faces, tetrahedral arrangement and different coordination numbers (CN), namely the CN12, CN14, CN15 and CN16 polyhedron. Therefore, phases with this structural arrangement are referred to as "Frank-Kasper phases", which include the A15, Laves (C15, C14, C36), σ- and μ-phase [9]. Shoemaker and Shoemaker [10] were the first to report that the interstices of these tetrahedrally close-packed phases are also exclusively tetrahedral. Today, these phases are more commonly referred to as topologically close-packed (TCP) phases, that form one of the largest groups of intermetallics [10–12]. The Laves phases structure is based on the stacking of two structural units: one is a single layer of smaller B atoms in a Kagomé net and the other one is a triple layer that can be described as an A-B-A sandwich like structure unit, consisting of two layers of the larger A atoms with a layer of B atoms in between [13]. For hexagonal Laves phases, the basal planes are formed by these layers [11, 13]. The μ-phase with the ideal stoichiometry of $A_6B_7$, and $W_6Fe_7$ as its prototype structure, exhibits a rhombohedral crystal structure containing 13 atoms per unit cell [14, 15]. It can be characterised as an intergrowth of two different structures, the $Zr_4Al_3$ and $MgCu_2$ structure, that are alternately stacked on top of each other and share the B atoms Kagomé net [16, 17].

Intermetallics are brittle at room temperature [12, 18]. The brittle-to-ductile transition temperature (BDTT) is at around 0.6 $T_m$ for Laves phases [18, 19]. Krämer and Schulze [13] studied the influence of the complex TCP structure, resulting in a high Peierls barrier to dislocation motion [20], on the plasticity of Laves phases. To investigate the plasticity of such brittle materials at room temperature, which requires the suppression of cracking, nanomechanical testing is a suitable method, by which the mechanical properties can also be determined [21]. By analysing the formed slip steps around indents, the influence of the phase, composition and grain orientation on the active slip systems can be studied within different small single grains, whereby inhomogeneities are also not considered as a criterion for exclusion [21–23].

Common slip systems in hexagonal crystals are basal, prismatic 1st and 2nd order (will be referred to as prismatic I and II in the following), as well as pyramidal 1st and 2nd order (will be referred to as pyramidal I and II in the following) [24]. Those slip systems have also been reported in literature for the C14 Laves phase [13, 25, 26]. For the μ-phase, in which the Laves phase ($MgCu_2$) structure is stacked alternately with the Zr$_4$Al$_3$ structure [17], plastic deformation can potentially appear in both structural units. However, previous studies on the Mo$_6$Fe$_7$ µ-phase [27] and on different stoichiometric and off-stoichiometric µ-phase compositions in the Nb-Co system [28, 29] concluded that the Laves phase layer is the main deformation carrier. Despite the common slip systems, with the basal plane identified as the favoured slip plane [28, 27, 29], non-basal slip along $\{1\ \bar{1}\ 0\ 5\}\ \langle \bar{5}\ 5\ 0\ 2\rangle$ was also observed, as reported in our previous work [30].

An important deformation mechanism that cannot go unmentioned in connection with Laves phases or Laves phase building blocks is the synchroshear mechanism. Hazzledine et al. [31–33] published their theory on this mechanism, which involves the synchronous movement of two Shockley partial dislocations within the triple layer of the Laves phase, based on previous findings of Kronberg [34] on sapphire. One of these two Shockley partial dislocations crosses the closely spaced A and B atom interlayers of the triple layer between the lower A atom and middle B atom interlayer and the other one between the middle B atom and upper A atom interlayer [31, 33, 35]. By moving two Shockley partial dislocations with a Burgers vector of $b = \frac{1}{3} \langle 1\ 0\ \bar{1}\ 0 \rangle$ each, a total displacement along the Burgers vector $b = \frac{1}{3} \langle 1\ 1\ \bar{2}\ 0 \rangle$ is achieved [31]. Atomistic simulations, supported by ab initio calculations, confirmed the synchroshear mechanism to be the most favourable mechanism for basal slip in Laves phases [36].

In this study, we focus on the Ta-Fe and Ta-Fe-Al systems, both containing a hexagonal C14 Laves and a µ-phase [37–39]. For Ta, Fe and Al atoms, the metallic radii are 146, 126 and 143 pm, respectively [40, 41]. Therefore, in the binary system the Ta atoms sit on the larger A and the Fe atoms on the smaller B sites, resulting in the stoichiometric TaFe$_2$ and Ta$_6$Fe$_7$ phases [37–39]. When a third element is added to the binary system, either the sites of the A or B atoms can be occupied. It was found that Al usually replaces the B component, although when comparing the metallic radii it becomes apparent that the atomic size of the alloyed Al is closer to that of Ta as A component whose substitution is also possible [40–42]. Due to the extended homogeneity ranges for the Laves and µ-phase in these systems [37–39], the investigation of off-stoichiometric sample compositions is also feasible and we consider such compositions here to study how the mechanical properties change within the phase depending on the composition. For the Laves phase structure, it was found that the extension to either the A- or B-rich side leads to point defects such as anti-site atoms or vacancies [18, 43–45]. If the µ-phase is not stoichiometric in composition due to an extension to the A-rich side, the B atom sites in the triple layer are more favourable for substitution by A atoms than the sites in the Kagomé net that consists exclusively of B atoms [46, 47, 28]. Accordingly, the triple layer of the other stoichiometric composition of the µ-phase, here Ta$_7$Fe$_6$, consists exclusively of the larger A (Ta) atoms [28].

The aim of this work is to reveal the mechanical properties and activated slip systems across Laves and µ-phases systematically. To this end, we selected the binary Ta-Fe and ternary Ta-Fe(-Al) systems, in which the Laves and µ-phase both exhibit a wide homogeneity range, allowing a systematic investigation of the mechanical properties and activated slip systems depending on the phase, composition and grain orientation. Using nanoindentation and analysing the resulting slip lines around the indents, we investigate the mechanical properties (hardness and elastic modulus) as well as active slip planes. The results of the slip trace analysis are confirmed using transmission electron microscopy (TEM) and the experimental moduli compared with first principles calculations. Based on this work, we aim to contribute to a better understanding of defects and deformation in the Laves and µ-phase in a well-controlled compositional space in order to inform further studies of their role in the deformation of high performance structural alloys and our ability to control plasticity in brittle TCP phases in functional materials.

## 2    Experimental methods

### 2.1    Sample synthesis and metallographic preparation

All samples were arc melted from the elements using input materials with a purity of at least 99.99 %, as described in detail in our previous work [48]. Different compositions close to the stoichiometry as well as off-stoichiometric were prepared in the Ta-Fe system, resulting in three different compositions for the binary Laves phase (28, 33 and 38 at.% Ta; rest Fe) and four for the binary µ-phase (46, 50, 54 and 58 at.% Ta; rest Fe). These samples are referred to in the following as $\lambda_{b1}$, $\lambda_{b2}$ and $\lambda_{b3}$ or $\mu_{b1}$, $\mu_{b2}$, $\mu_{b3}$ and $\mu_{b4}$ for the Laves or µ-phase, respectively. In the Ta-Fe-Al system, the Ta content was kept constant at 33 at.% for the Laves phase and at 54 at.% for the µ-phase and only the Fe:Al ratio was varied. Three different compositions were prepared for the ternary Laves phase (33 at.% Ta; 22, 33.5 and 45 at.% Al; rest Fe) and ternary µ-phase (54 at.% Ta; 9, 16 and 23 at.% Al; rest Fe), respectively. In the following they are referred to as $\lambda_{t1}$, $\lambda_{t2}$ and $\lambda_{t3}$ (Laves phase) or $\mu_{t1}$, $\mu_{t2}$ and $\mu_{t3}$ (µ-phase). All sample names, their system, phase and target composition are also given in Table II in section 3.1.1.

One binary Ta-Fe and one ternary Ta-Fe-Al µ-phase sample were used for preparing TEM lamellae. The binary sample has the same target composition as sample $\mu_{b3}$ (54 at.% Ta) but was prepared with a different Ta as input material (purity > 99.9 %). The ternary µ-phase sample has a slightly different composition than sample $\mu_{t3}$ with the Ta content being reduced by 2 at.% Ta (52 at.% Ta) but an equal Fe:Al ratio of 1:1 (24 at.% each). As input material, a semi-finished Ta with a basis purity of 99.95 % was used. The Fe and Al raw materials are the same as for the 13 samples introduced first. These two samples were also prepared by arc melting. Reasons for the different input materials are optimisations of the microstructure by using raw materials with different degrees of purity and adjusting the synthesis route. They were explained in detail in our previous work [48].

For the metallographic preparation of the samples, they were first ground with increasingly fine grit using SiC paper (500 to 4000 grit size) and then semi-automatically polished (diamond suspension with 3 µm and 1 µm particle size) for several hours. Finally, the samples were polished with an oxide polishing suspension (OP-U). To perform the subsequent nanomechanical experiments, the samples were mounted on SEM stubs with Crystalbond 509 (Aremco) due to the high stiffness of the adhesive. The transition of the sample edge towards the stub was brushed with silver glue to ensure good conductivity.

## 2.2　Chemical and microstructural characterisation

To verify the synthesised phases and corresponding compositions electron backscatter diffraction (EBSD) and energy-dispersive X-ray spectroscopy (EDS) measurements were performed (Helios NanoLab 600i from FEI Inc. equipped with Hikari XP2 EBSD detector and Octane Super A EDS detector, both EDAX). The EBSD investigations for microstructural analysis were conducted with an acceleration voltage of 20 kV and a beam current between 2.7 and 5.5 nA, on a 70° pre-tilted sample holder, at a working distance between 10 and 11.5 mm. The grain orientations of the inverse pole figure (IPF) refer to the sample normal. For the chemical analysis by EDS an acceleration voltage of 20 kV and a beam current of 2.7 nA were chosen. The measurements were performed at 4 mm working distance. More details on the evaluation of these measurements can be found in our previous work [48], where the successful preparation of the targeted phases with compositions close to the targeted values was confirmed.

A Python-based automation tool [49] integrated in the SEM (CLARA from TESCAN) was used to find indent positions in a pre-defined area. The SE images were captured in ultra-high resolution (UHR) mode. An accelerating voltage of 5 kV and a beam current of 30 pA at a working distance of 1.2 to 1.3 mm were chosen.

## 2.3　Nanomechanical characterisation and evaluation

Nanoindentation tests were performed at room temperature using a load-controlled system (iNano Nanoindenter from KLA Instruments) equipped with a diamond Berkovich indenter tip (supplied by Synton-MDP). The continuous stiffness measurement (CSM) was used and the analysis was performed according to the method defined by Oliver and Pharr [50, 51] to measure the hardness and elastic modulus. In this work, the elastic modulus determined by indentation is referred to as the indentation modulus [52], which already includes the correction of the deformation of the indentation tip calculated on the basis of the reduced modulus [50] as well as the complex stress state of indentation, which can be classified between uniaxial and hydrostatic. The experiments were carried out with a maximum load of 45 mN at a strain rate of 0.2 $s^{-1}$ resulting in a final depth between 350 and 400 nm depending on the tested sample. For calculating the hardness and indentation modulus, the data from 250 nm indentation depth until the end of the test were used. A constant Poisson ratio of 0.32 [53] was used for all samples and the elastic constants for diamond as indenter tip material (E = 1140 GPa, v = 0.07) [50, 51].

On each of the 13 different samples at least one standard indentation array with 100 indents was performed. A fused silica sample was tested as reference (4 x 4 indents) before performing the indentation tests on each sample to calibrate the area function of the tip. Preliminary investigations had shown that due to the high hardness of the studied intermetallic phases, a strong tip wear can be observed. The extent of this may depend on the previous wear of the tip but was considerable here. If this wear is not considered by recalibration and the area function therefore only inadequately describes the actual tip geometry when analysing the tests, the measured mechanical properties are not correct and the comparison of the different samples loses its validity.

How significantly the area function changed after testing one sample can be seen in Figure S1 in section S1 in the supplementary material.

By correlating the slip lines around the indents on the sample surface, captured by SE imaging, with the orientation information from the EBSD measurements, a slip trace analysis could be carried out. Its purpose is to determine the activated slip planes and the activation frequency of these depending on the grain orientation and the tested sample – system, phase and composition. By fabricating special sample holders, misalignment between the sample positions for EBSD and SE measurements were minimised as they allow the same indent orientations due to a unique position of the SEM stub during both acquisitions. However, only the intersection of the activated plane with the sample surface is visible, i.e. only two-dimensional information are available, which is subject to remaining uncertainty from surface and sample alignment in the SEM. For this reason, a threshold angle of 3° was defined and all planes within the thus defined range were included as possible slip planes to ensure that no potential planes are neglected. This results in a total activation frequency of more than 100 %. A more detailed description of this method can be found in literature [23]. The recorded SE images were also used to determine if the indents can be included in the analysis. Reasons for excluding individual indents were the presence of a second phase, that was partially or entirely tested and testing in or near an impurity, pore or visible surface scratches.

For the slip trace analysis on the binary and ternary Laves and µ-phase of the Ta-Fe(-Al) system, the primary slip systems for hexagonal metals [24] were examined. In addition, the $\{1\,\bar{1}\,0\,5\}$ plane observed in our previous work [30] and the $\{1\,\bar{1}\,0\,26\}$ A-B-A triple layer plane found in TEM were also included in the analysis of the µ-phase. Thereby the A-B-A notation applies to the prototype stoichiometric µ-phase and the position within the unit cell is highlighted in the inset of Figure 7 (b) for the binary and in Figure 9 (b) for a ternary µ-phase. Furthermore, it has to be mentioned that the last index of the Miller-Bravais indices is dependent on the c/a ratio and therefore varies slightly for different µ-phases. All analysed slip systems for the Laves and µ-phase are listed in Table I with the number of equivalent planes of the slip system and its resulting theoretical frequency.

Table I: Possible slip planes for the C14 Laves phase and additional systems for the µ-phase given with the Miller-Bravais indices of the slip systems. The number of equivalent planes per slip system and their frequency (rounded) when considering all possible slip planes per phase are also included.

| Phase | Slip Plane | Miller-Bravais Indices (Slip System) | # Equivalent Planes | | Frequency [%] | |
|---|---|---|---|---|---|---|
| | | | Laves | µ | Laves | µ |
| Laves | basal | $(0\,0\,0\,1)\langle 1\,1\,\bar{2}\,0\rangle$ | 1 | 1 | 5.3 | 4.5 |
| | prismatic I | $\{1\,0\,\bar{1}\,0\}\langle 1\,1\,\bar{2}\,0\rangle$ | 3 | 3 | 15.8 | 13.6 |
| | prismatic II | $\{1\,1\,\bar{2}\,0\}\langle 1\,\bar{1}\,0\,0\rangle$ | 3 | 3 | 15.8 | 13.6 |
| | pyramidal I | $\{1\,0\,\bar{1}\,1\}\langle 1\,1\,\bar{2}\,3\rangle$ | 6 | 3 | 31.6 | 13.6 |
| | pyramidal II | $\{1\,1\,\bar{2}\,2\}\langle 1\,1\,\bar{2}\,3\rangle$ | 6 | 6 | 31.6 | 27.3 |
| + µ | $\{1\,\bar{1}\,0\,5\}$ | $\{1\,\bar{1}\,0\,5\}\langle \bar{5}\,5\,0\,2\rangle$ | - | 3 | - | 13.6 |
| | $\{1\,\bar{1}\,0\,26\}$ | $\{1\,\bar{1}\,0\,26\}\langle 1\,1\,\bar{2}\,0\rangle$ | - | 3 | - | 13.6 |

## 2.4 TEM investigation

The microstructures around two indents, one in a binary Ta-Fe and another one in a ternary Ta-Fe-Al µ-phase sample were examined and analysed by TEM. The grain orientation given by the Euler angles (Bunge) are $(302.5, 37.7, 80.6)$ for the binary and $(79.5, 39.1, 286.7)$ for the ternary sample, both favouring the activation of basal slip. Therefore, the lamellae positions were chosen perpendicular to the basal plane within the centre of the individual indents and prepared by focused ion beam (FIB) milling to achieve a uniform sample thickness. Initially, prior to the milling process, a thin Pt layer was deposited on the region of interest to minimise beam damage. Subsequently, a Ga$^+$ ion source was utilised to mill the lamellae at an acceleration voltage of 30 kV and a beam current between 21 nA and 80 pA, which was decreased with decreasing lamella thickness. Final polishing of the lamellae was performed at a lower acceleration voltage of 5 kV and a beam current of 41 pA. Due to the stress field within the TEM sample, resulting from the plastic deformation by indentation, strong bending of the lamellae was expected and minimised by attaching it on both sides to the Cu-grid. The dimensions of the thinned TEM sample area are about 4 µm in width and 3.5 µm in height with a lamella thickness of approximately 75 nm for the binary Ta-Fe µ-phase. It has a hole in the bottom region of the thinned area due to bending of the lamella. The TEM sample of the ternary Ta-Fe-Al µ-phase has a thickness of approximately 70 nm and a width and height of approximately 4 by 5 µm. A thickness of less than 100 nm was aimed for because of the high coordination number of Ta (Z = 73) [41]. The characterisation of the indented microstructures was carried out in a conventional TEM (JEM-F200 from JEOL) at an acceleration voltage of 200 kV. The microscope is equipped with a double-tilt holder that enables an X-tilt of ± 36° and a Y-tilt of ± 31°.

## 2.5 Computational modelling

Electronic structure calculations were conducted using Density Functional Theory (DFT) in the Vienna A*b Initio* Software Package (VASP) [54, 55]. Planewave basis sets, represented by the Projector Augmented Wave (PAW) potential (2015 release), were used to describe the systems one-electron orbital wavefunctions [56]. Only the valence states were considered bonding states, which are Fe $3d^7$ $4s^1$, Ta $5p^6$ $5d^4$ $6s^1$ and Al $3s^2$ $3p^1$. A kinetic cut-off energy of 550 eV to the planewave basis was applied. The Perdew, Burke and Ernzerhof (PBE) exchange-correlation functional was used within the generalised gradient approximation (GGA) level of theory [57]. A smearing of 0.05 at the Fermi level was done using the method of Methfessel-Paxton of order 1. Electronic and geometric optimisation convergence criteria were set to $10^{-6}$ eV and 0.02 eV/Å respectively.

The site occupancy of Al in the C14 Ta-Fe-Al Laves phase was previously explored [58]. The stability of eight structural motifs, each corresponding to different Fe:Al ratios in $Ta_4Fe_{8-x}Al_x$ (x = 0 to 7), were investigated in a formation energy phase diagram. It was found that only the $Ta_4Fe_6Al_2$ and $Ta_4Fe_2Al_6$ show large stability regions, indicating that Al has strong preference to be spaced out among the Kagomé layers before reaching saturation ($Ta_4Fe_2Al_6$). In this work, the properties of $Ta_4Fe_6Al_2$, $Ta_4Fe_3Al_5$ and $Ta_4Fe_2Al_6$ were included to show the effect of a systematic increase in Al content.

The lattice parameters of the C14 TaFe$_2$, Ta$_4$Fe$_6$Al$_2$, Ta$_4$Fe$_3$Al$_5$ and Ta$_4$Fe$_2$Al$_6$ Laves phases and the Ta$_6$Fe$_7$ and Ta$_7$Fe$_6$ µ-phases were optimised in DFT by firstly scanning the energy along the two-dimensional a and c parameters, preserving the α, β and γ angles. Based on the energy minima, a full-cell relaxation scheme of each cell was further implemented to obtain a more accurate determination of the cell lattice parameters. Upon the substitution of Al into the C14 Ta-Fe Laves phase structure, the hexagonal lattice symmetry was broken. Gamma-cantered k-point meshes of 10×10×5 and 5×5×1 were applied on the C14 Laves and µ-phase conventional unit cell structures respectively.

To calculate the elastic tensor, the energy-strain approach was used, where energy can be expressed as a Taylor expansion of strain, described in Equation (1):

$$E(V, \{\varepsilon_j\}) = E(V_0, 0) + V_0 \sum_{i}^{6} \sigma_i \varepsilon_i + \frac{V_0}{2} \sum_{i,j=1}^{6} c_{ij} \varepsilon_i \varepsilon_j + \cdots \quad (1)$$

where $E(V_0, 0)$ and $V_0$ are the energy and volume of the equilibrium structure. The second-order energy derivative with respect to strain was used to derive the elastic tensor. In this approach, distortion modes based on the crystal symmetry were applied to the equilibrium structure and strain energies were calculated for each structure. The AELAS package [59] was used to automatically generate the distorted structures and fit the energy-distortion relationship to determine the quadratic coefficients and the elastic tensor components. At least 1000 k-points per reciprocal atom (KPPRA) were assigned in all structures when determining the new gamma-cantered k-point mesh. Given that crystal symmetry changes with composition, we have provided all the elastic constants ($C_{11}$, $C_{12}$, $C_{13}$, $C_{23}$, $C_{22}$, $C_{33}$, $C_{44}$, $C_{55}$, $C_{66}$) for an elastically orthotropic material.

The Young's modulus was used to evaluate the average tensile stiffness of a material, where a lower value corresponds to a lower stiffness. In this work, we used Hill's ($E_{VRH}$) approximation [60] of the Young's modulus, which is an arithmetic average of Voigt's ($K_V$, $G_V$, $E_V$) and Reuss's ($K_R$, $G_R$, $E_R$) definitions [61, 62], i.e. $K_{VRH} = ½ (K_V + K_R)$ and $G_{VRH} = ½(G_V + G_R)$. All equations are shown below:

$$9K_V = (c_{11} + c_{22} + c_{33}) + 2(c_{12} + c_{23} + c_{31}) \quad (2)$$
$$15G_V = (c_{11} + c_{22} + c_{33}) - (c_{12} + c_{23} + c_{31}) + 4(c_{44} + c_{55} + c_{66})$$

$$\frac{1}{K_R} = (s_{11} + s_{22} + s_{33}) + 2(s_{12} + s_{23} + s_{31}) \quad (3)$$
$$\frac{15}{G_R} = 4(s_{11} + s_{22} + s_{33}) - 4(s_{12} + s_{23} + s_{31}) + 3(s_{44} + s_{55} + s_{66})$$

$$E = \frac{9KG}{3K + G} \tag{4}$$

As the magnetic configurations of Ta-Fe and Ta-Fe-Al Laves phases have been systematically explored in a previous work [58], the lowest-energy configurations for $TaFe_2$, $Ta_4Fe_6Al_2$, $Ta_4Fe_3Al_5$ and $Ta_4Fe_2Al_6$ were used for our comparative analysis. The antiferromagnetic (AF) configuration was deduced as the lowest-energy configuration in DFT, which aligns with experimental observations [58, 63]. Due to the paramagnetic property of Fe at room temperature, we also provide the second and third-lowest energy configurations ($TaFe_{2(Fi)}$ for the ferrimagnetic and $TaFe_{2(Fo)}$ for the ferromagnetic ordering, respectively), to give insight into the effect of magnetic ordering on the Young's modulus.

Similarly, several magnetic configurations within the µ-phase are energetically close to each other, which can result in some variability of the elastic properties. DFT exploration of the magnetic configurations shows that ferromagnetic ordering of the Fe atoms in the stoichiometric µ-phase results in the lowest energy configuration. To highlight the influence of atom occupancy on the elastic stiffness of the crystal, the elastic properties of $Ta_6Fe_7$ and $Ta_7Fe_6$ with identically ferromagnetic Fe were calculated. This comparison assumes that the substitution of Ta does not change the magnetic ordering of Fe.

## 3  Results

### 3.1  Hardness and elastic modulus

In Figure 1, the average hardness and indentation modulus values given with their standard deviations in GPa are plotted in (a) against the target Ta content in at.% for the seven binary TCP phase samples and in (b) against the target Al content in at.% for the six ternary ones, where the Ta content was kept constant at 33 at.% Ta for the ternary Laves and at 54 at.% Ta for the ternary µ-phase samples. Additionally, the Young's modulus for different compositions in the Ta-Fe(-Al) system as calculated by DFT is included in Figure 1. The elastic moduli of antiferromagnetic, ferromagnetic and ferrimagnetic $TaFe_2$ Laves phases are shown to provide insight into the effect of magnetic configurations. Upon comparing the effect of composition within the binary µ-phase and ternary Laves phase, the lowest-energy magnetic configurations were antiferromagnetically-ordered Fe and ferromagnetically-ordered Fe, respectively.

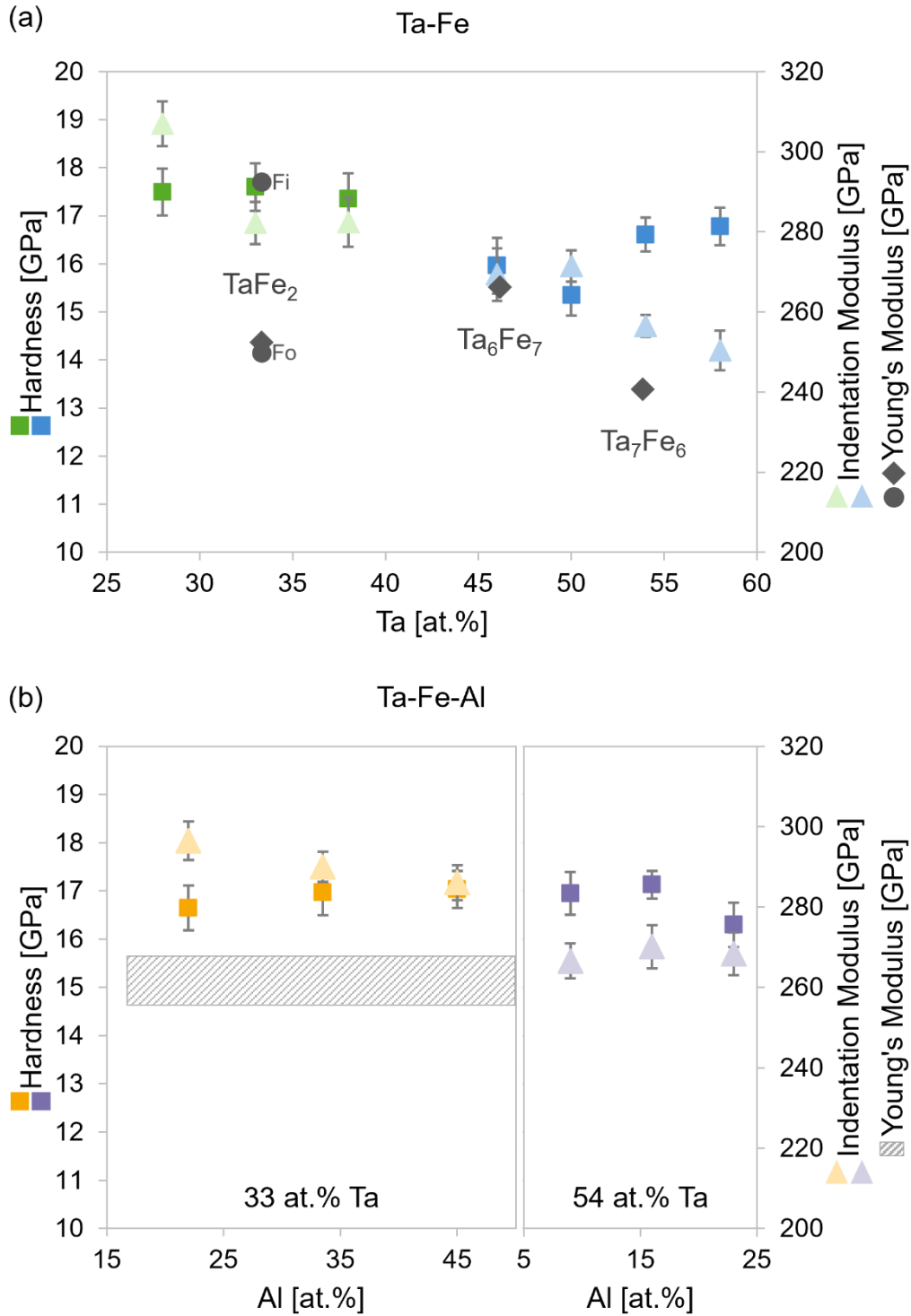

*Figure 1: Average hardness and indentation modulus (given in GPa) of the (a) binary Ta-Fe samples plotted against the target Ta content (in at.%) and the (b) ternary Ta-Fe-Al samples plotted against the target Al content (in at.%). The measured values are given with their calculated standard deviations. Squares of a more intense colour represent the hardness values while the values of the indentation modulus are illustrated by triangles in a translucent colour. The binary Laves and µ-phase are coloured green and blue, respectively, while the ternary phases are shown in orange (Laves phase) and purple (µ-phase). The Young's modulus of the lowest-energy configuration predicted by DFT (given in GPa) is also included for $TaFe_{2(AF)}$, $Ta_6Fe_7$ and $Ta_7Fe_6$ and represented by the dark grey squares. The Laves phase predictions with further alternative magnetic ordering, $TaFe_{2(Fi)}$ and $TaFe_{2(Fo)}$, are marked by the dark grey circles. The Young's modulus range of the three theoretically stable ternary Laves phase compositions, $Ta_4Fe_6Al_2$, $Ta_4Fe_3Al_5$ and $Ta_4Fe_2Al_6$, is highlighted by the light grey hatched bar.*

3.1.1   Nanoindentation

All experimentally determined hardness and indentation modulus values that are illustrated in Figure 1 are also listed in Table II together with their standard deviations. In addition, this table contains information on the phase, sample name and composition (as described in section 2.1) as well as on the number of indents per sample that were actually used to determine the mechanical properties. After excluding indents that are within a second phase, on a phase boundary, within an impurity or on top of a deeper scratch, a minimum of 60 tests were included per sample. The sample with the lowest number of included indents is sample $\mu_{b1}$ with 46 at.% Ta, in which the Laves phase is present as a second phase within the µ-phase matrix due to its composition at the phase field boundary in the phase diagram, cf. [37]. In total, 1108 tests were included in the analysis shown in Figure 1 and listed in Table II.

*Table II: Average hardness and indentation modulus given with their standard deviation in GPa for the different Laves and µ-phase samples of the binary and ternary Ta-Fe(-Al) system. The samples are listed with their target compositions of Ta, Fe and Al in at.%. The number of indents used to determine the mechanical properties per sample is also included.*

| System | Phase | Sample | Target composition [at.%] | | | Hardness [GPa] | Indentation Modulus [GPa] | # Indents |
|---|---|---|---|---|---|---|---|---|
| | | | Ta | Fe | Al | | | |
| Ta-Fe | Laves | $\lambda_{b1}$ | 28 | 72 | - | 17.5 ± 0.5 | 307 ± 6 | 61 |
| | | $\lambda_{b2}$ | 33 | 67 | - | 17.6 ± 0.5 | 282 ± 5 | 96 |
| | | $\lambda_{b3}$ | 38 | 62 | - | 17.4 ± 0.5 | 282 ± 6 | 97 |
| | µ | $\mu_{b1}$ | 46 | 54 | - | 16.0 ± 0.6 | 269 ± 7 | 60 |
| | | $\mu_{b2}$ | 50 | 50 | - | 15.4 ± 0.4 | 271 ± 4 | 95 |
| | | $\mu_{b3}$ | 54 | 46 | - | 16.6 ± 0.4 | 257 ± 3 | 99 |
| | | $\mu_{b4}$ | 58 | 42 | - | 16.8 ± 0.4 | 250 ± 5 | 99 |
| Ta-Fe-Al | Laves | $\lambda_{t1}$ | 33 | 45 | 22 | 16.7 ± 0.5 | 297 ± 5 | 91 |
| | | $\lambda_{t2}$ | 33 | 33.5 | 33.5 | 17.0 ± 0.5 | 290 ± 4 | 80 |
| | | $\lambda_{t3}$ | 33 | 22 | 45 | 17.0 ± 0.4 | 286 ± 4 | 97 |
| | µ | $\mu_{t1}$ | 54 | 37 | 9 | 17.0 ± 0.4 | 267 ± 4 | 100 |
| | | $\mu_{t2}$ | 54 | 30 | 16 | 17.1 ± 0.3 | 270 ± 5 | 67 |
| | | $\mu_{t3}$ | 54 | 23 | 23 | 16.3 ± 0.5 | 268 ± 5 | 66 |

The hardness values for the binary Laves phase samples are in a narrow range between approximately 17.4 and 17.6 GPa. However, within the standard deviation the nearly stoichiometric sample exhibits the highest hardness. A strong difference can be observed for the indentation modulus that reaches its maximum at 307 GPa for sample $\lambda_{b1}$ with 28 at.% Ta and therefore the highest Fe content. The other two binary Laves phase samples ($\lambda_{b2}$ and $\lambda_{b3}$) have a similar indentation modulus of around 282 GPa. In case of the binary µ-phase samples, the indentation modulus is lower for all samples than that of the Laves phase samples. While the first two µ-phase samples ($\mu_{b1}$ and $\mu_{b2}$) show a similar indentation modulus of around 270 GPa, this value drops to approximately 250 GPa for sample $\mu_{b4}$ with the highest Ta content of 58 at.% Ta. In between, the indentation modulus of sample $\mu_{b3}$ (54 at.% Ta) was measured at around 257 GPa. The hardness values show an opposite trend with increasing Ta content, at first slightly decreasing and then increasing again. The lowest value was measured for sample $\mu_{b2}$ at 15.4 GPa and the highest for sample $\mu_{b4}$ at 16.8 GPa. Within the ternary system, in which the Fe:Al ratio is varied, the Laves phase samples exhibit a similar hardness in the range of approximately 16.7 to 17.0 GPa. The indentation modulus shows a slightly decreasing trend from 297 GPa (sample $\lambda_{t1}$) to 286 GPa (sample $\lambda_{t3}$) with increasing Al content. For the ternary μ-phase samples, a lower indentation modulus in the range of 267 to 270 GPa was measured. For sample $\mu_{t1}$ and $\mu_{t2}$, with 9 and 16 at.% Al, respectively, the hardness value is at around 17.0 GPa and then slightly decreases to 16.3 GPa for sample $\mu_{t3}$ with the highest Al content. However, for sample $\mu_{t3}$ it has to be mentioned that the orientation of the μ-phase as sample matrix could not clearly be assigned which might influence the determined mechanical properties. The standard deviations for the measured hardness and indentation modulus values of all samples are low in this analysis suggesting a consistent data set, cf. Table I. The maximum values were calculated for sample $\mu_{b1}$ (46 at.% Ta) with a standard deviation of 3.8 % for the hardness and 2.6 % for the indentation modulus. Most indents were excluded for samples $\lambda_{b1}$, $\mu_{b1}$, $\mu_{t2}$ and $\mu_{t3}$ that exhibit a (finely dispersed) second phase due to their composition being close to the boundary of the phase filed.

### 3.1.2  Density functional theory

The Young's modulus of various compositions in the Laves and μ-phases was calculated in DFT to validate the decreasing indentation modulus trend between the Laves phase and the μ-phases. The stoichiometric Laves phase (33.33 at.% Ta) and the stoichiometric prototype μ-phase (46.15 at.% Ta) were modeled by $TaFe_2$ and $Ta_6Fe_7$, respectively. The other stoichiometric μ-phase (53.84 at.% Ta) was modeled as the $Ta_7Fe_6$ structure, where Ta substitutes all Fe atoms on the triple-layer sites.

All DFT lattice and elastic moduli for $TaFe_2$, $Ta_4Fe_6Al_2$, $Ta_4Fe_3Al_5$, $Ta_4Fe_2Al_6$, $Ta_6Fe_7$ and $Ta_7Fe_6$ are listed in Table III. $TaFe_{2(AF)}$, $TaFe_{2(Fo)}$ and $TaFe_{2(Fi)}$ have different magnetic configurations, where $TaFe_{2(AF)}$ is the lowest-energy structure among the three. Depending on the magnetic configuration, lattice parameters can vary between 0.2 to 0.5 %, but are in overall good agreement with experimental data for both the Laves [42, 64–69] and the μ-phase [70]. Our calculations show that the substitution of Fe with Al in the ternary Laves phase and with Ta on the triple layer B site of the μ-phase ($Ta_7Fe_6$) both expand the lattice.

Within the building blocks of the μ-phase, the increased Ta content leads to a shift in the distances between the basal interlayers. In particular, the interplanar distances associated with the slip within the Laves building block increase, consistent with a previous work on isostructural Nb-Co μ-phases [28]. We consider the distances in the Laves phase building block between the middle and lower levels of the triple layer ($d_t$) and between the lower level of the triple layer and the Kagomé layer ($d_{t-K}$), as well as in the $Zr_4Al_3$ building block, specifically the distances between the Kagomé layer and the upper A level ($d_{K-CN14}$), and between the upper and middle A level ($d_{CN14-CN15}$) (these are illustrated in detail in [28]). Upon forming $Ta_7Fe_6$ in the μ-phase, $d_t$, the active interlayer for crystallographic basal slip, increases from 0.369 to 0.382 Å, and $d_{t-K}$ increases from 1.637 to 1.730 Å. Conversely, the spacings $d_{K-CN14}$ and $d_{CN14-CN15}$, decrease from 1.136 to 1.069 Å and 1.361 to 1.333 Å, respectively.

*Table III: Lattice parameters, total magnetic moment per formula unit (μ$_B$/f.u.) and Young's modulus (E) of DFT-simulated TaFe$_2$, Ta$_4$Fe$_6$Al$_2$, Ta$_4$Fe$_3$Al$_5$, Ta$_4$Fe$_2$Al$_6$, Ta$_6$Fe$_7$ and Ta$_7$Fe$_6$. The Young's modulus is given in Voigt, Reuss and Hill's approximation. TaFe$_{2(AF)}$, TaFe$_{2(Fo)}$ and TaFe$_{2(Fi)}$ have different magnetic configurations, with TaFe$_{2(AF)}$ being the lowest-energy structure among the three. The Wyckoff label for each unique atomic position is detailed under atomic positions.*

| Parameter | TaFe$_2$ (AF) | TaFe$_2$ (Fo) | TaFe$_2$ (Fi) | Ta$_4$Fe$_6$Al$_2$ | Ta$_4$Fe$_3$Al$_5$ | Ta$_4$Fe$_2$Al$_6$ | Ta$_6$Fe$_7$ | Ta$_7$Fe$_6$ |
|---|---|---|---|---|---|---|---|---|
| a [Å] | 4.776 | 4.784 | 4.757 | 4.861 | 4.963 | 4.982 | 4.855 | 4.939 |
| b [Å] | 4.776 | 4.784 | 4.757 | 4.861 | 4.983 | 5.025 | 4.855 | 4.939 |
| c [Å] | 7.880 | 7.843 | 7.861 | 7.982 | 8.178 | 8.245 | 26.831 | 27.097 |
| μ$_B$/f.u. | 0.8 | 2.2 | 0.5 | 0.0 | -1.5 | 0.0 | 9.0 | 7.6 |
| E (Voigt) [GPa] | 255 | 261 | 295 | 267 | 269 | 269 | 270 | 244 |
| E (Reuss) [GPa] | 250 | 239 | 290 | 245 | 264 | 266 | 263 | 237 |
| E (Hill) [GPa] | 252 | 250 | 292 | 256 | 267 | 268 | 266 | 241 |
| Atomic Positions | (2a)Fe (4f)Ta (6h)Fe | (2a)Fe (4f)Ta (6h)Fe | (2a)Fe (4f)Ta (6h)Fe | (2a)Al (4f)Ta (6h)Fe | (2a)Al (4f)Ta (6h)Al, Fe | (2a)Al (4f)Ta (6h)Al, Fe | (3a)Fe (6c)Ta (18h)Fe | (3a)Ta (6c)Ta (18h)Fe |

In terms of the Young's modulus, a strong dependence on magnetic ordering is observed (Figure 1). TaFe$_{2(Fi)}$ and TaFe$_{2(AF)}$, the lowest energy structures, are nearly isoenergetic, with an energy difference of ~0.007 eV, while TaFe$_{2(Fo)}$ is approximately 0.02 eV higher. While the Young's modulus comparison between TaFe$_{2(Fi)}$ and the μ-phase qualitatively aligns with the experimental observation, the trend changes when compared to TaFe$_{2(AF)}$. The reason for this is unclear. An in-depth study on the magnetic ordering in the μ-phase would be required to resolve this issue, similar to our previous work on the smaller unit cell of the Laves phase [58], which is beyond the scope of this predominately experimental study.

Among the ternary Ta-Fe-Al compositions explored in the Laves phase, Ta$_4$Fe$_2$Al$_6$ models a saturated Ta-Fe-Al composition of 33.33 at.% Ta, 16.78 at.% Fe, and 50 at.% Al, while Ta$_4$Fe$_6$Al$_2$ and Ta$_4$Fe$_3$Al$_5$ simulate two other theoretically stable ternary Laves phases [58]. At finite temperatures, the Ta-Fe-Al ordering is expected to consist of a distribution of stable and metastable structural motifs, making a direct one-to-one indentation modulus comparison with DFT-calculated moduli impractical. We therefore give in Figure 1 the range of the Young's modulus between the stable states, where Ta$_4$Fe$_6$Al$_2$ and Ta$_4$Fe$_2$Al$_6$ present the minimum and maximum at 256 and 268 GPa, respectively. Given that DFT-calculated elastic moduli usually have a margin error of approximately 2 %, this range of 12 GPa is considered narrow and the effect of Al therefore small to negligible. Thus, Fe can be substituted by Al within the Ta-Fe Laves phase at approximately constant stiffness, which is in line with the experimental observations.

## 3.2 Slip trace analysis

### 3.2.1 Orientation dependence of the slip trace morphology

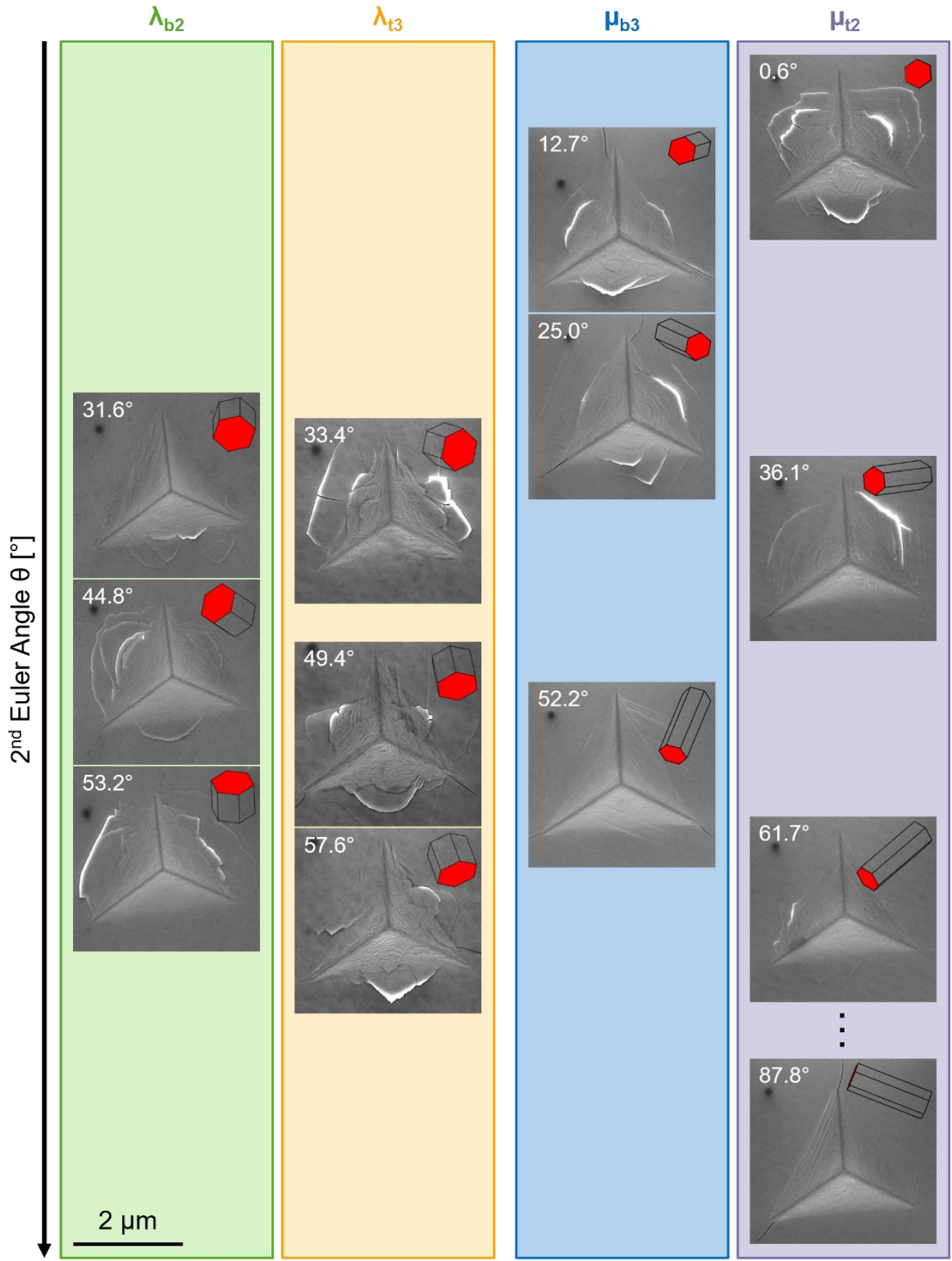

*Figure 2: Typical indents of the binary Laves phase sample λ$_{b2}$ with 33 at.% Ta, the ternary Laves phase λ$_{t3}$ with 33 at.% Ta and 45 at.% Al, the binary μ-phase sample μ$_{b3}$ with 54 at.% Ta and the ternary μ-phase sample μ$_{t2}$ with 54 at.% Ta and 16 at.% Al sorted by the second Euler angle θ (given in °). All samples contain Fe as the binary or ternary element to balance the alloys to 100 at.%. The unit cells are inset with the basal plane being highlighted in red.*

Typical indents of four different samples, one of each system and phase together with the oriented unit cell and its basal plane highlighted in red, are sorted by the second Euler angle θ as shown in Figure 2. The nearly stoichiometric binary Laves phase sample $\lambda_{b2}$ with a target composition of 33 at.% Ta is given in the first column and the ternary Laves phase sample $\lambda_{t3}$ with the same Ta content and the highest Al amount of 45 at.% is placed next to it in the second column. Besides the two Laves phase samples, two µ-phase samples are also included in column three and four in Figure 2. For the binary Ta-Fe system, indents of sample $\mu_{b3}$ are shown, since this sample with its composition of 54 at.% Ta can be best compared to the ternary $\mu_{t2}$ sample with the same Ta content and 16 at.% Al. It can clearly be seen how the morphology of the representative indents changes for the different phases, compositions and orientations. A categorisation of the different types of slip lines surrounding indents has already been defined in previous publications [27, 71, 72]. Thereby a distinction was made between edges (segments of different slip planes), lines (straight, only individual planes), curved (curvature of slip lines not allowing the assignment to the possible slip planes), only cracks, few surface features or no visible surface traces.

The indents of the binary Laves phase $\lambda_{b2}$ show only a few surface traces mostly of the edge type for θ of about 30°. As θ increases, curved slip traces can mainly be observed for θ nearly 45° until a mixture of the edge type slip traces, in which longer straight lines as well as round segments become present for θ equal to 53.2°. Similar findings can be obtained for the ternary Laves phase sample $\lambda_{t3}$, where indents in grains of comparable orientations could be analysed. Although the indents for θ close to 30° already show more surface lines in the form of edges and straight segments, there is also a tendency towards curved slip lines with increasing θ (at approximately 50°) and then again, a mixture of edges with longer straight and curved segments at θ equal to 57.6°.

In case of the binary µ-phase $\mu_{b3}$, for the orientation with low θ, mainly slip lines of the curved type can be recognised, which become straight with increasing inclination of the basal plane towards 45°. Already for θ equal to 25° slip lines of the edge type as well as some straight and parallel lines can be observed. These parallel lines seem to be assignable to the basal plane. For orientations where the Schmid factor for the basal plane is high, only straight, parallel lines that mainly belong to the activated basal plane can be identified (see representative indent for θ equal to 52.2°). However, the activation of non-basal slip planes is to be expected as well in the plastic zone below the indent due to the complex stress field resulting from indentation. To form a visible slip step at the sample surface, a large number of dislocations is necessary. We find that this is achieved for the µ-phase on the basal plane, but not on the non-basal planes. For the ternary µ-phase sample $\mu_{t2}$, clear slip lines can be recognised for the orientation in which the basal plane is almost parallel to the sample surface (θ close to 0°) that consist of many segments and therefore correspond mainly to the edge type. With increasing θ, the amount of slip lines of the edge type decreases and, as for the binary µ-phase, more fine, straight and parallel surface traces become visible, seemingly associated to the basal plane. If the inclination of the basal plane reaches a position almost perpendicular to the sample surface, only a few slip lines are still visible and primarily cracks can be observed. For both µ-phases, the binary and the ternary one, it can be seen that the fewest cracks are observed for orientations with θ close to 45°, whereby many parallel slip lines are present due to the activation of the basal plane.

### 3.2.2 Activated slip planes

The slip trace analysis was performed for different grain orientations of 12 different samples of the Ta-Fe(-Al) system. Sample µ$_{t3}$, the µ-phase sample with an equal Fe:Al ratio, could not be included due to the already mentioned ambiguous assignability of the orientation of the µ-phase as sample matrix. In total, 459 indents were included within the performed slip trace analysis, resulting in 4994 analysed, sufficiently straight and long slip lines. The threshold angle between the identified slip line in the SE image and the calculated intersection line of the activated slip plane and the sample surface from orientation information is defined as 3°. All possible slip planes within this angle were counted, resulting in 6900 possibly activated slip planes and, therefore, in a total activation frequency of approximately 138 % for all analysed samples.

In the IPFs shown in Figure 3 and Figure 4 for the binary Ta-Fe Laves and µ-phase samples and in Figure 5 and Figure 6 for the TCP phases of the ternary Ta-Fe-Al system, the relative slip plane activation for different examined slip planes is given within the IPF legend considering all three Euler angles ($\varphi_1, \theta, \varphi_2$) (Bunge). For the IPFs only the relative activation frequency, i.e. the proportional distribution of the activated slip planes, adding up to 100%, can be represented. The actual activation frequency for all tested orientations is given in the supplementary material in section S2, where for each sample and each orientation the activation frequency is plotted over the second Euler angle θ that directly correlates with the inclination of the basal plane. Per grain orientation, ten indents were analysed. Orientations for which fewer than ten indents could be analysed due to the grain size or the position of the indent array are, however, also included in the IPFs in order to achieve the most complete dataset possible. They are illustrated transparently and have a dashed or dotted frame, depending on whether the data is available for five to nine or less than five indents, respectively. In the following, the five analysed slip planes for the Laves phase samples, the (0 0 0 1) basal, {1 0 $\bar{1}$ 0} prismatic I and {1 1 $\bar{2}$ 0} prismatic II as well as the {1 0 $\bar{1}$ 1} pyramidal I and {1 1 $\bar{2}$ 2} pyramidal II planes, are coloured in red, blue, turquoise, green and yellow, respectively. Additionally, the {1 $\bar{1}$ 0 5} plane and the {1 $\bar{1}$ 0 26} plane were included in the slip trace analysis for the µ-phase samples and are coloured in magenta and white, respectively.

#### 3.2.2.1 Binary Ta-Fe system

In the binary Ta-Fe system, all seven samples, three Laves and four µ-phase samples with different Ta contents, have been analysed.

*Laves phase*

For the binary Laves phase samples, results displayed in Figure 3, 141 indents with 1642 slip traces (1931 slip traces including multiple counts within the threshold angle of 3°) were examined, observing mainly non-basal slip. Due to the texture of these samples prepared by arc melting, predominantly orientations near the centre of the IPF with the second Euler angle close to 45° could be tested. Because of the restrictions in sample preparation [48], it was not possible to reprepare textured samples to increase the orientation spread. For the binary Ta-Fe Laves phase samples the share of basal slip is low for under stoichiometric and stoichiometric TaFe$_2$, as illustrated in Figure 3 (a) and (b) but increases significantly for sample λ$_{b3}$ with 38 at.% Ta shown in Figure 3 (c). When considering the binary Laves phase samples in total, the $\{1\,1\,\bar{2}\,2\}$ pyramidal II plane is the most frequently encountered slip plane, followed by the $\{1\,0\,\bar{1}\,1\}$ pyramidal I plane, while the $\{1\,1\,\bar{2}\,0\}$ prismatic II plane was the rarest found slip plane. However, especially for sample λ$_{b1}$ and λ$_{b2}$ (Figure 3 (a) and (b)), it can be observed that the amount of prismatic slip traces significantly increases for orientations closer to the right, i.e. the $[1\,0\,\bar{1}\,0]$ and $[2\,\bar{1}\,\bar{1}\,0]$ directions, of the IPF.

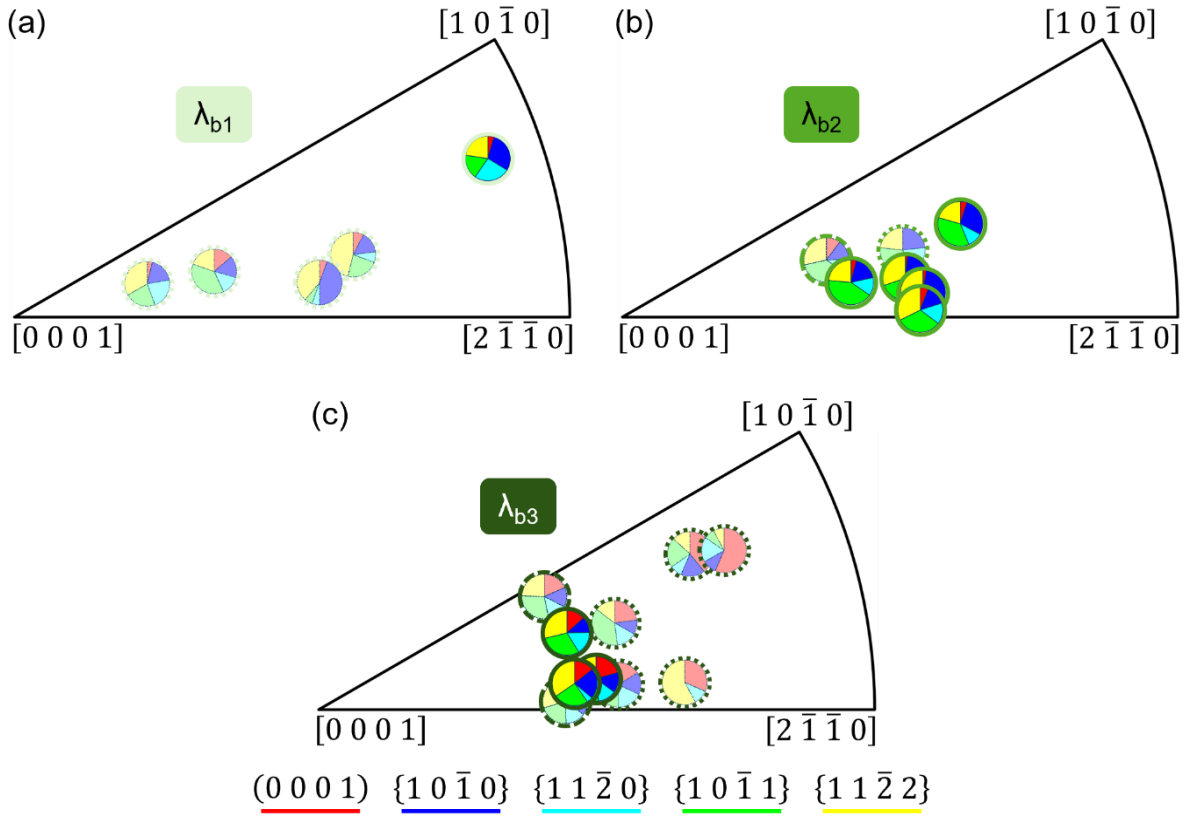

Figure 3: Relative activation frequency of the different slip planes illustrated as pie chart for the analysed orientations within the IPF of the binary Laves phase sample λ$_{b1}$ with 28 at.% Ta in (a), λ$_{b2}$ with 33 at.% Ta in (b) and λ$_{b3}$ with 38 at.% Ta in (c). Fe balances all samples to 100 at.%. The colour scheme for the different slip planes is given. Orientations for which ten indents could be examined are displayed in bright colours with continuous frames, while orientation where only five to nine or less than five indents could be analysed are highlighted slightly transparent with a dashed or dotted frame, respectively.

*μ-phase*

To study the activated slip planes in the binary μ-phase, 166 indents of the four different compositions and 1525 slip traces (with 2278 solutions for these slip traces) were examined. The results of this analysis are given in Figure 4 for each binary μ-phase composition. It is immediately noticeable that the proportion of basal slip is overall significantly higher than for the Laves phase, cf. Figure 3. Sample μ$_{b3}$ in Figure 4 (c), for which orientations in the centre of the IPF and therefore with θ close to 45° could be tested in particular shows the large proportion of basal slip. Besides the basal slip plane, comparatively large shares of the $\{1\,\bar{1}\,0\,5\}$ and $\{1\,\bar{1}\,0\,26\}$ planes can be identified, especially for orientations near the right side of the IPF as for sample μ$_{b1}$ in Figure 4 (a) close to the $[1\,\bar{1}\,0\,0]$ direction. Sample μ$_{b4}$, the μ-phase sample with the highest Ta content of 58 at.% Ta (Figure 4 (d)) exhibits a higher proportion of non-basal slip. Even for orientations with θ close to 45° in the IPF centre, where the Schmid factor for basal slip reaches its maximum, the relative activation of the basal plane does not exceed one third for this composition. Furthermore, samples µ$_{b3}$ and µ$_{b4}$ (Figure 4 (c) and (d)) show that basal slip is hardly detected for orientations close to the [0 0 0 1] direction, where θ is near 0° (low Schmid factor and low visibility of any slip traces extending approximately parallel to the surface) and increases with increasing inclination angle of the basal plane towards the IPF centre. The total activation frequency of the different slip planes is plotted over the second Euler angle θ for the binary µ-phase compositions in Figure S3 (section S2 of the supplementary material). On average, it is of the order of 150 % for the binary µ-phase samples, which is due to double indexation, especially for the basal and high-indexed pyramidal planes. Overall, the basal plane is dominant in spite of three or six times fewer equivalent planes for potential double indexing, followed by the $\{1\,\bar{1}\,0\,26\}$ plane. Slip lines indicating deformation along the $\{1\,0\,\bar{1}\,0\}$ prismatic I plane were found least for the orientations analysed.

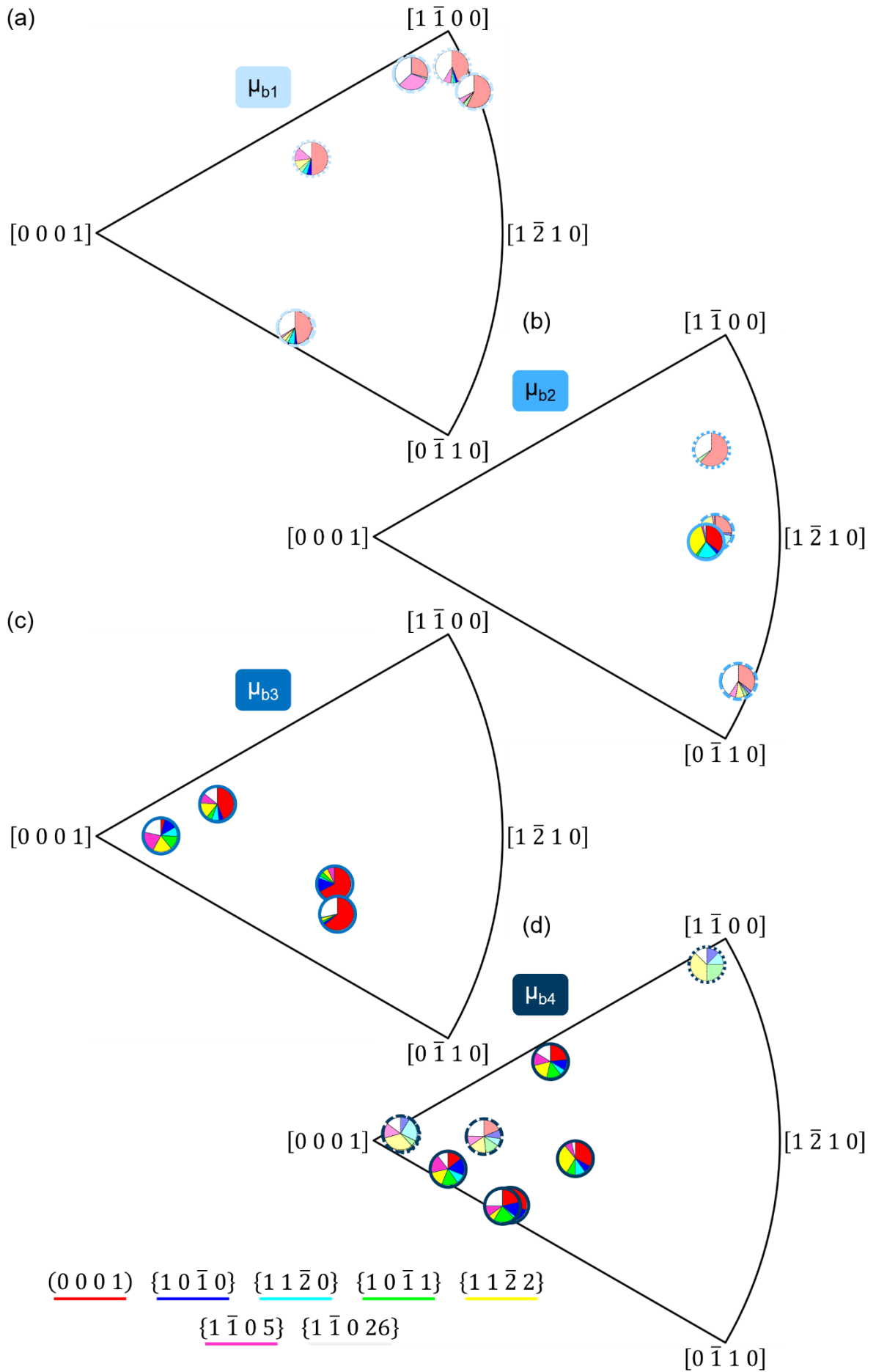

*Figure 4: Relative activation frequency of the different slip planes illustrated as pie chart for the analysed orientations within the IPF of the binary µ-phase sample µ$_{b1}$ with 46 at.% Ta in (a), µ$_{b2}$ with 50 at.% Ta in (b), µ$_{b3}$ with 54 at.% Ta in (c) and µ$_{b4}$ with 58 at.% Ta in (d). Fe balances all samples to 100 at.%. The colour scheme for the different slip planes is given. Orientations for which ten indents could be examined are displayed in bright colours with continuous frames, while orientation where only five to nine or less than five indents could be analysed are highlighted slightly transparent with a dashed or dotted frame, respectively.*

### 3.2.2.2 Ternary Ta-Fe-Al system

In total, five samples were analysed for the ternary Ta-Fe-Al system, three Laves phase and two µ-phase samples that exhibit different Fe:Al ratios at constant Ta contents.

*Laves phase*

For the Laves phase samples, 65 indents with 816 slip traces (1002 slip traces including multiple solutions) were analysed, and the relative slip plane activation for different grains within the IPF is given for all three sample compositions in Figure 5. Due to the texture of these samples and the large grain size, the number of orientations that could be tested was unfortunately lower than for the other phases. For example, for sample $\lambda_{t2}$ in Figure 5 (b), only one orientation was included due to the large grain size. As for the binary Laves phase compositions, orientations near the IPF centre with θ close to 45° could be primarily observed and tested. Plastic deformation seems to occur primarily by slip along non-basal planes. The $\{1\,1\,\bar{2}\,2\}$ pyramidal II plane, closely followed by the $\{1\,0\,\bar{1}\,1\}$ pyramidal I plane and the $\{1\,0\,\bar{1}\,0\}$ prismatic I plane, are the in total most frequently observed slip planes for the analysed compositions and orientations of the ternary Laves phase. The basal and $\{1\,1\,\bar{2}\,0\}$ prismatic II planes are the second least and least observed slip planes, respectively.

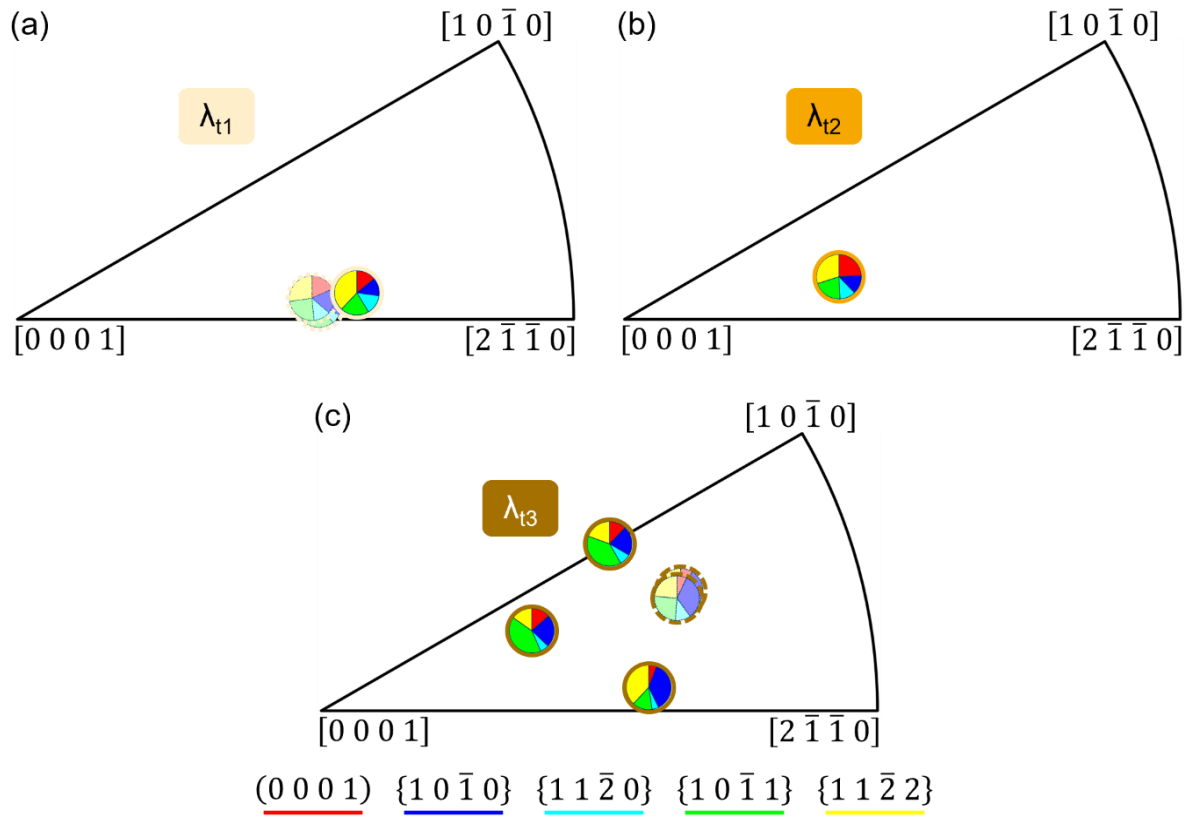

Figure 5: Relative activation frequency of the different slip planes illustrated as pie chart for the analysed orientations within the IPF of the ternary Laves phase sample λ$_{t1}$ with 45 at.% Fe and 22 at.% Al in (a), λ$_{t2}$ with 33.5 at.% Fe and 33.5 at.% Al in (b) and λ$_{t3}$ with 22 at.% Fe and 45 at.% Al in (c). All samples have a constant Ta content of 33 at.%. The colour scheme for the different slip planes is given. Orientations for which ten indents could be examined are displayed in bright colours with continuous frames, while orientation where only five to nine or less than five indents could be analysed are highlighted slightly transparent with a dashed or dotted frame, respectively.

*µ-phase*

The slip trace analysis for the two remaining ternary µ-phase samples µ$_{t1}$ and µ$_{t2}$ was performed on 87 indents and 1011 slip traces (1689 slip trace solutions within the 3° threshold angle). The results are displayed in Figure 6. The proportion of basal slip is overall lower than in the binary µ-phase while slip along the $\{1\,1\,\bar{2}\,2\}$ pyramidal II plane is most frequently observed for the tested sample areas. However, given the lower number of equivalent planes for the basal plane, it is still identified as likely the one with the lowest critical resolved shear stress (CRSS). In terms of relative activation, the $\{1\,\bar{1}\,0\,26\}$ plane was determined to be the third most frequent slip plane and the $\{1\,0\,\bar{1}\,0\}$ prismatic I plane the least frequent one. The total activation frequency is on average 163 %.

For sample µ$_{t1}$ in Figure 6 (a), it is clearly visible how the share of the basal plane activation increases with the increase of θ towards 45°, while the activation of the basal plane was hardly observed for orientations close to the $[0\,0\,0\,1]$ direction. As for the binary µ-phase sample µ$_{b1}$ (Figure 4 (a)), for sample µ$_{t2}$ in Figure 6 (b) and the tested orientation close to the $[1\,\bar{1}\,0\,0]$ direction, slip along the basal and $\{1\,\bar{1}\,0\,26\}$ planes was mainly identified. But even in the centre of the IPF with θ equal to 62°, the relative activation of the basal plane for the tested orientation of sample µ$_{t2}$ is less than one quarter. The orientation closest to the $[1\,\bar{2}\,1\,0]$ direction shows a relatively high proportion of prismatic II and pyramidal II slip, however, for this orientation, these planes are close to the basal plane as slip trace and are therefore probably double counted.

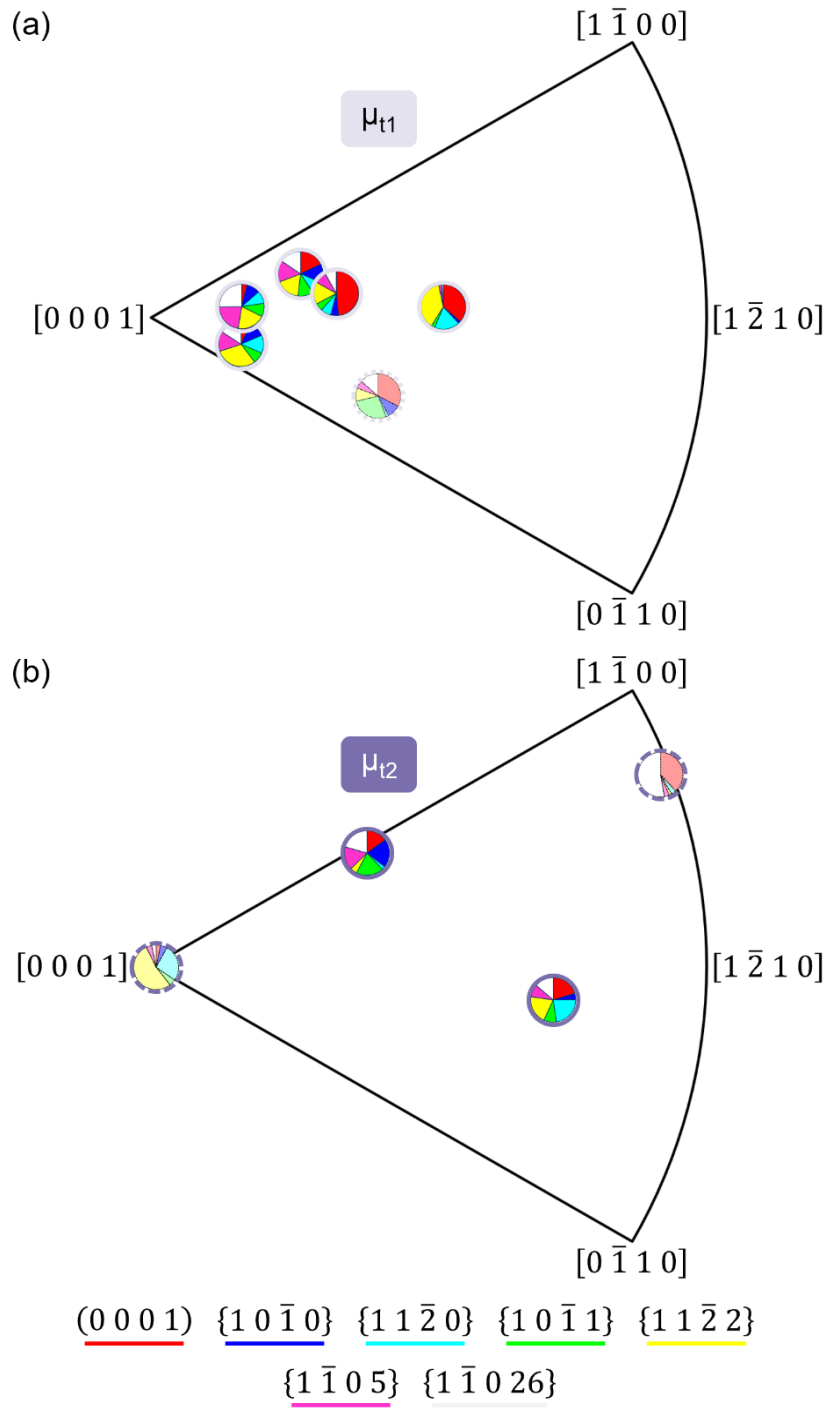

Figure 6: Relative activation frequency of the different slip planes illustrated as pie chart for the analysed orientations within the IPF of the ternary µ-phase sample µ$_{t1}$ with 37 at.% Fe and 9 at.% Al in (a) and µ$_{t2}$ with 30 at.% Fe and 16 at.% Al in (b). Both samples have a constant Ta content of 54 at.%. The colour scheme for the different slip planes is given. Orientations for which ten indents could be examined are displayed in bright colours with continuous frames, while orientation where only five to nine or less than five indents could be analysed are highlighted slightly transparent with a dashed or dotted frame, respectively.

## 3.3 TEM

### 3.3.1 Binary Ta₇Fe₆ μ-phase

Two TEM lamellae of a binary and ternary μ-phase were investigated. A typical indent of the binary Ta-Fe μ-phase with 54 at.% Ta (rest Fe) is shown in Figure 7 (a) as SEM SE image. To study the plastic zone underneath it in a TEM, a lamella was prepared from the indent, see TEM bright field (BF) image in Figure 7 (b). Several different slip bands (indicated by the letters "A" to "F") can be observed underneath the indent (Figure 7 (b)). The TEM BF image was taken at zone axis $[2\,\bar{1}\,\bar{1}\,0]$, in which three slip bands are edge-on. They are on the $(0\,0\,0\,1)$ basal plane (slip band A, indicated by the red line), on the $(0\,1\,\bar{1}\,5)$ plane (slip band C, marked by the magenta line) and one slip band is close to the $(0\,1\,\bar{1}\,26)$ A-B-A triple layer plane (slip band B, indicated by the white line). Additionally, the non-edge-on slip bands D, E and F can be observed. According to the TEM image, slip band E appears to have different segments, while slip band F is far away from edge-on. The white dashed square reveals the area where slip bands D, E and F were further analysed at other zone axes given in Figure 8. Cracks on the basal planes are also observed under the indent and are marked by red arrows. Some other cracks do not appear to be located along the basal plane. However, it could not be clearly determined along which plane they are situated

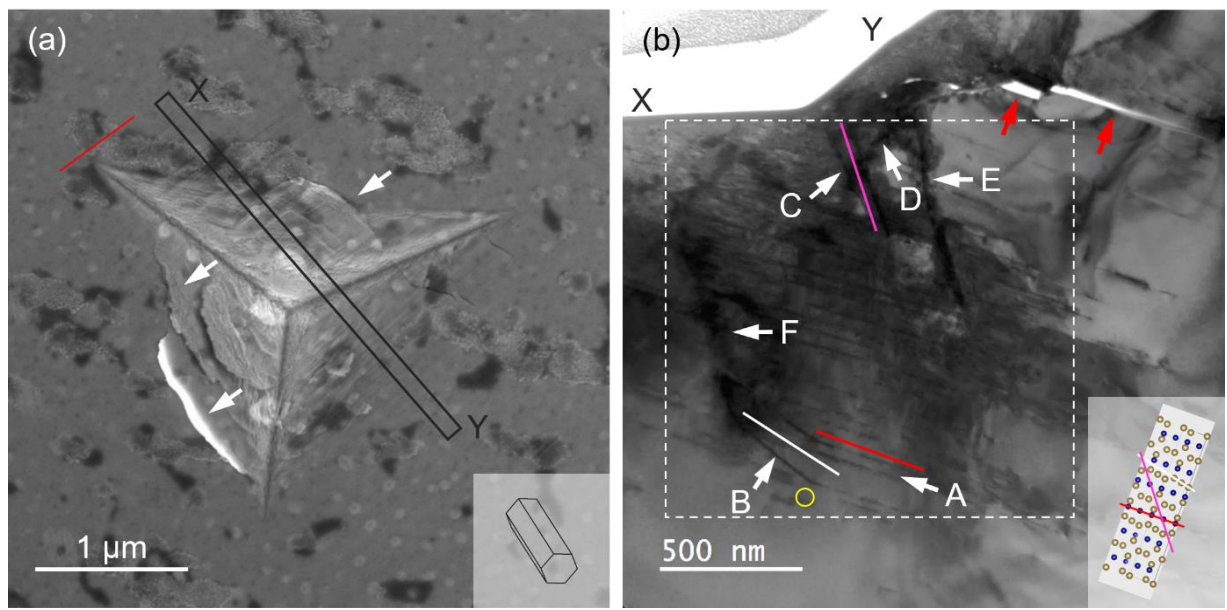

Figure 7: (a) SEM SE image of a typical indent of the Ta-Fe μ-phase with 54 at.% Ta. The unit cell is inset and the intersection of the basal plane with the sample surface is highlighted as basal slip trace (red line). Other wavy slip traces around the indent are indicated by the white arrows. The position of the lamella is marked as a dark grey rectangle and defined by the letters "X" and "Y". (b) TEM BF image of the microstructure underneath the indent taken at zone axis $[2\,\bar{1}\,\bar{1}\,0]$, in which six slip bands were observed (indicated by the letters "A" to "F"). The unit cell of the binary Ta₇Fe₆ μ-phase is inset, Ta atoms are coloured in gold and Fe atoms in blue (visualisation using VESTA [73]). The yellow circle shows the location where the selected area diffraction (SAD) aperture was placed.

To determine the slip planes of slip bands D, E, and F, the lamella was further tilted to zone axes $[1\,0\,\bar{1}\,0]$ and $[10\,0\,\overline{10}\,1]$. The TEM BF images taken at zone axes $[2\,\bar{1}\,\bar{1}\,0]$, $[1\,0\,\bar{1}\,0]$ and $[10\,0\,\overline{10}\,1]$ and the corresponding selected area diffraction (SAD) patterns are shown in Figure 8 (a), (b) and (c), respectively. Slip bands C and D are determined to be on the $(0\,1\,\bar{1}\,5)$ and $(1\,\bar{1}\,0\,2)$ plane, respectively. Another slip band that appears to be nearly parallel to slip band D in the top left corner seems to be on the $(2\,\bar{1}\,\bar{1}\,2)$ plane. Additionally, slip band E is observed to have different sections and was therefore indexed to be on different planes. While its top section is on the $(1\,0\,\bar{1}\,1)$ or $(\bar{2}\,1\,1\,2)$ plane, the lower section could be identified to be on the $(0\,1\,\bar{1}\,5)$ plane. Besides that, is noticed that the lower section of slip band E is sheared by basal dislocations (magenta arrows in (b)). Figure 8 (d) shows the magnified morphology of slip band C and E taken at a high index zone axis close to zone axis $[1\,0\,\bar{1}\,0]$. In this image it can be seen that both slip bands C as well as the lower section of slip band E are sheared by basal stacking faults (indicated by the magenta arrows). A similar observation was also made in our previous work [30]. Due to the wide extension of slip band F at all the three zone axes reached in TEM, the slip plane could not be determined. Outside the plastic zone directly underneath the indent, prismatic slip was also observed within the binary µ-phase sample.

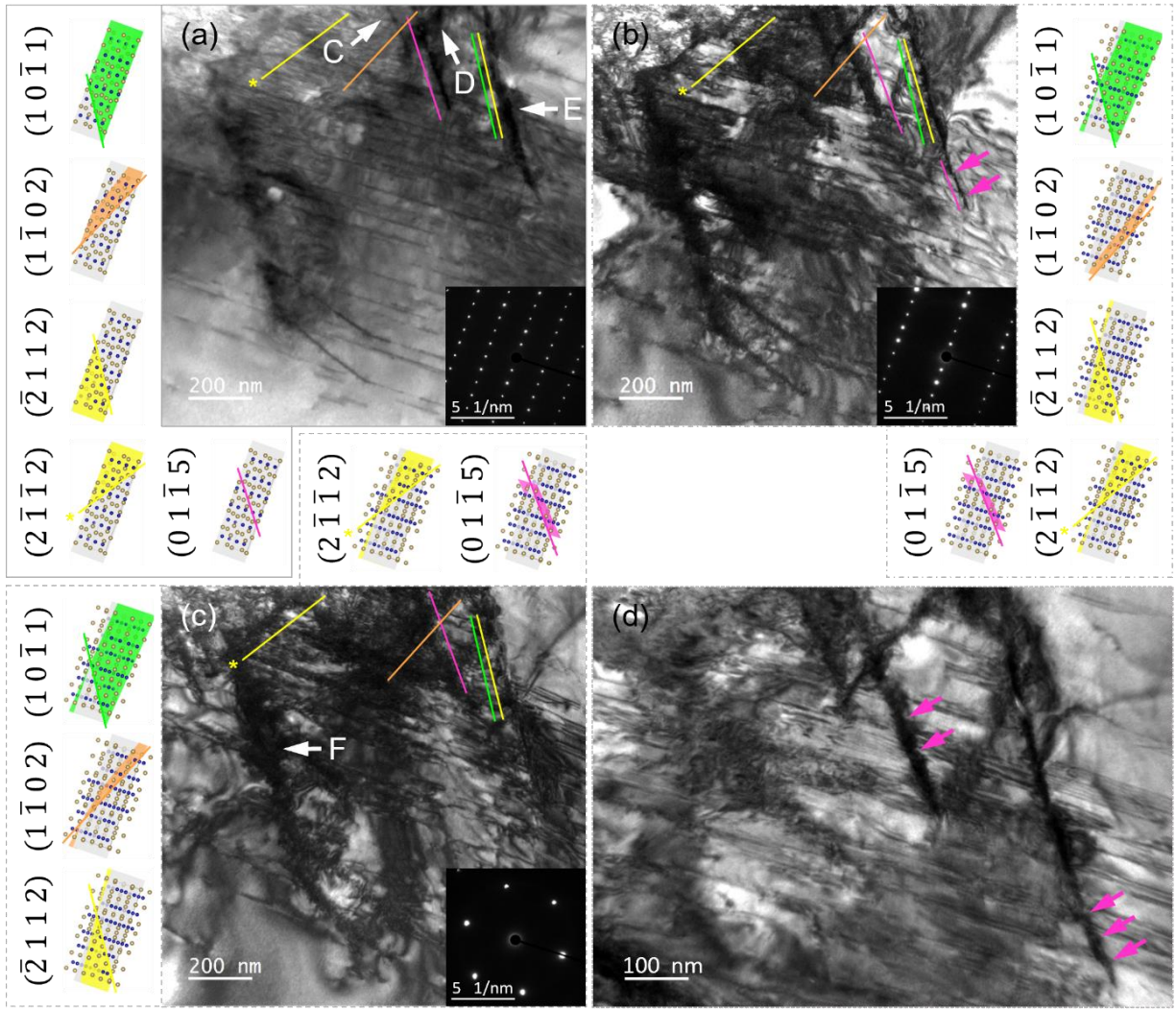

*Figure 8: TEM BF images taken at zone axis (a) $[2\,\bar{1}\,\bar{1}\,0]$, (b) $[1\,0\,\bar{1}\,0]$ and (c) $[10\,0\,\overline{10}\,1]$. The corresponding SAD patterns are inset. The unit cell of the binary Ta$_7$Fe$_6$ µ-phase is displayed next to the TEM images, Ta atoms are coloured in gold and Fe atoms in blue (visualisation using VESTA [73]). The intersecting lines of the approximated lamella (foil) plane $(39\,\overline{10}\,\overline{29}\,0)$ (highlighted in light grey) and the possible slip planes are provided within the unit cells. The intersecting lines are compared to the slip band morphology in the TEM images and can be determined accordingly. (d) Magnified TEM BF image taken at a high index zone axis close to zone axis $[1\,0\,\bar{1}\,0]$ showing both the slip band C and the lower section of slip band E are sheared by basal stacking faults (magenta arrows).*

Gasper et al. 2024    Mechanical properties and deformation mechanisms in Ta-Fe(-Al) TCP phases

### 3.3.2 Ternary µ-phase

The same analysis was carried out for the ternary Ta-Fe-Al µ-phase. A typical indent in the sample surface of the ternary Ta-Fe-Al µ-phase and its oriented unit cell are given in Figure 9 (a). A montage TEM BF image (consisting of several images at the same zone axis) of the milled lamella is shown in Figure 9 (b). It can be observed that the dislocation density right underneath the indent is high compared to the areas further away from the indent. Similar to the binary Ta-Fe µ-phase sample, several different slip bands can be observed in the lamella. These TEM images that form together the montage image were taken at zone axis $[\bar{1}\,\bar{1}\,2\,0]$ and the corresponding SAD pattern is provided in Figure 9 (c). The $(0\,0\,0\,1)$ basal plane (red line), the $(1\,\bar{1}\,0\,\bar{2})$ plane (orange line) and the $(1\,\bar{1}\,0\,26)$ plane (white line) are edge-on. Besides them, some other non-edge-on slip bands can be observed for which the identification of the activated slip plane is difficult. While the area where the SAD aperture was placed is exactly at zone axis $[\bar{1}\,\bar{1}\,2\,0]$, other areas slightly further away of it may be not fully in-zone due to the bending of the lamella. In order to more precisely determine the slip planes of the other observed slip bands, which are not edge-on at zone axis $[\bar{1}\,\bar{1}\,2\,0]$, further analysis was conducted. This analysis of the two areas marked by the dashed white squares in Figure 9 (b) are shown in Figure 10.

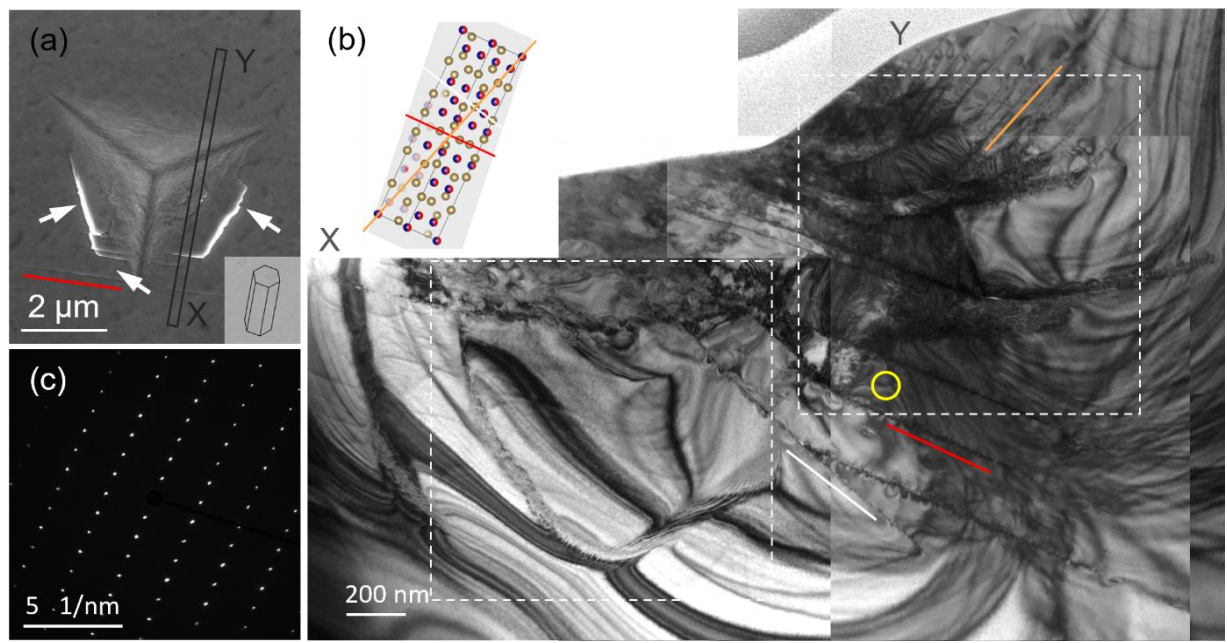

*Figure 9: (a) SEM SE image of a typical indent in the ternary Ta-Fe-Al µ-phase with 52 at.% Ta and 24 at.% Fe and Al each. The unit cell is inset and the intersection of the basal plane with the sample surface is highlighted as basal slip trace (red line). Other wavy slip traces around the indent are indicated by the white arrows. The position of the lamella is marked as a dark grey rectangle and defined by the letters "X" and "Y". (b) Montage TEM BF images providing an overview of the lamella, taken at zone axis $[\bar{1}\,\bar{1}\,2\,0]$. The yellow circle shows the location, where the SAD aperture was placed. The unit cell of the ternary Ta-Fe-Al µ-phase is inset, Ta atoms are coloured in gold, Fe atoms in blue and Al atoms in red (visualisation using VESTA [73]). (c) SAD pattern of the zone axis $[\bar{1}\,\bar{1}\,2\,0]$.*

To identify the other observed slip bands and verify our findings from the TEM BF image in Figure 9 (b) the SAD aperture was placed at different locations and the lamella was tilted to different zone axes due to the bending of the lamella. In Figure 10 (a) and (c) lamella was tilted to the $[\bar{1}\,\bar{1}\,2\,0]$ zone axis for two different areas, while the same areas were captured at the $[\bar{1}\,0\,1\,0]$ zone axis in Figure 10 (b) and (d), respectively. Possible slip planes for the µ-phase were included in the slip trace analysis. Four slip bands A,

B, C and D were determined by TEM. They could be identified as A: $(1\,0\,\bar{1}\,2)$, $(\bar{1}\,\bar{1}\,2\,2)$ or $(0\,1\,\bar{1}\,5)$ plane, B: $(1\,0\,\bar{1}\,\bar{2})$ plane, C: again $(1\,0\,\bar{1}\,2)$, $(\bar{1}\,\bar{1}\,2\,2)$ or $(0\,1\,\bar{1}\,5)$ plane and D: $(1\,0\,\bar{1}\,\bar{1})$ or $(0\,1\,\bar{1}\,5)$ plane. Additionally, another slip band in the top left corner that is comparable to slip band D was also analysed in Figure 10 (a) and (b) and identified as the $(1\,0\,\bar{1}\,\bar{1})$ or $(0\,1\,\bar{1}\,5)$ plane.

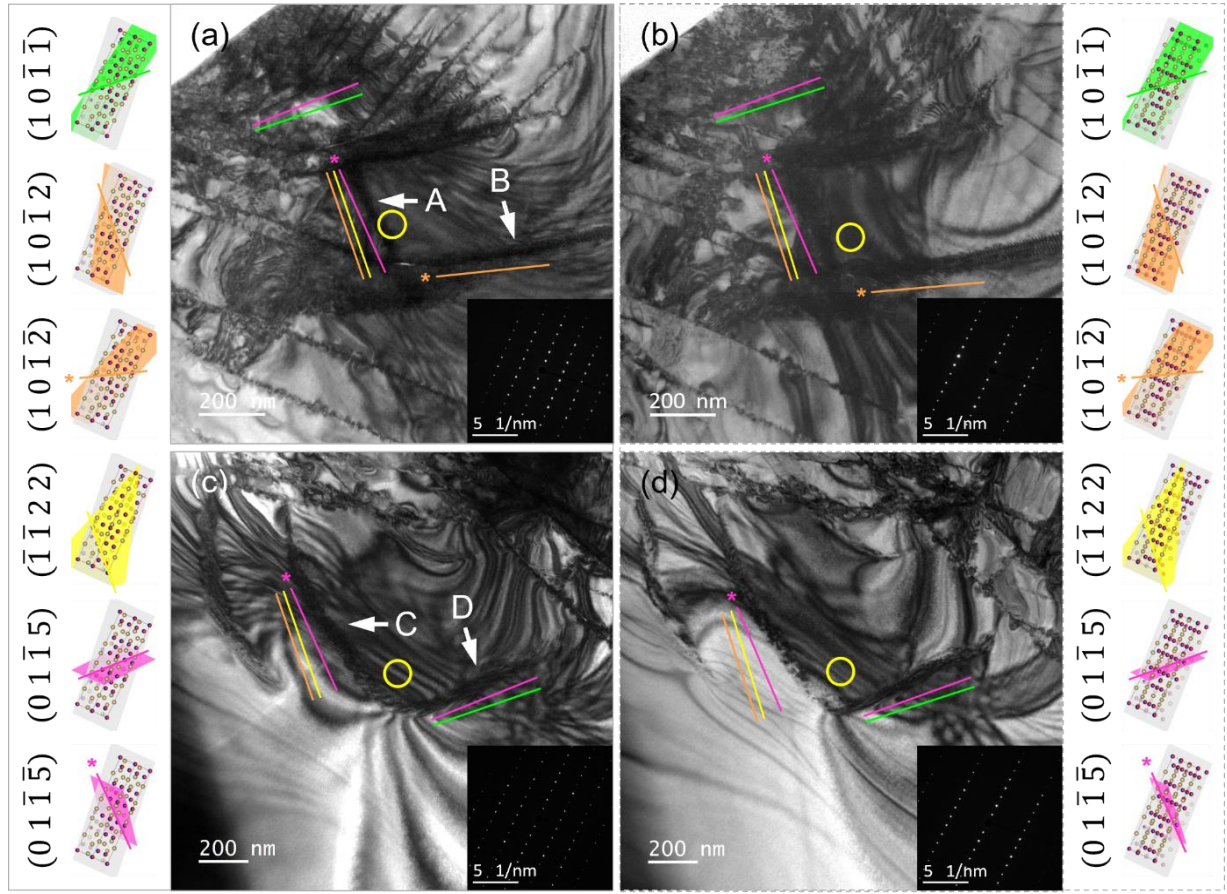

*Figure 10: (a) and (b) TEM BF images of the microstructure underneath the indent of the same area, taken at (a) $[\bar{1}\,\bar{1}\,2\,0]$ and (b) $[\bar{1}\,0\,1\,0]$ zone axes. (c) and (d), another pair of TEM BF images of a different area taken at (c) $[\bar{1}\,\bar{1}\,2\,0]$ and (d) $[\bar{1}\,0\,1\,0]$ zone axes. The corresponding SAD patterns are inset for all images and the yellow circles show the locations, where the SAD aperture was placed. The unit cell of the ternary Ta-Fe-Al µ-phase is displayed next to the TEM images, Ta atoms are coloured in gold, Fe atoms in blue and Al atoms in red (visualisation using VESTA [73]). The intersecting lines of the approximated lamella (foil) plane $(14\,5\,\overline{19}\,10)$ (highlighted in light grey) and the possible slip planes are compared to the slip band morphology in the TEM images. These lines are given in different colours and are positioned next to the observed slip bands for comparison. Accordingly, four slip bands A, B, C and D are determined.*

## 4   Discussion

### 4.1   Deformation mechanisms

Based on the large amount of generated data obtained from the high throughput but partly geometrically ambiguous nanoindentation slip trace analysis coupled with TEM on selected indentations, a number of new insights into the deformation mechanisms of the investigated TCP phases of the Ta-Fe(-Al) system were obtained. We present these separated first by crystal structure and the effect of chemistry within each, to discuss our insights into the activated slip systems particularly in the µ-phase, where these are poorly understood. In a second step, we compare the effect of composition on the relative activation of basal and non-basal mechanisms across the two crystal structures.

### 4.1.1 Laves phase

IPFs that contain all the different tested sample compositions of one TCP phase are shown in Figure 11 (a) for the binary and in (b) for the ternary Laves phases.

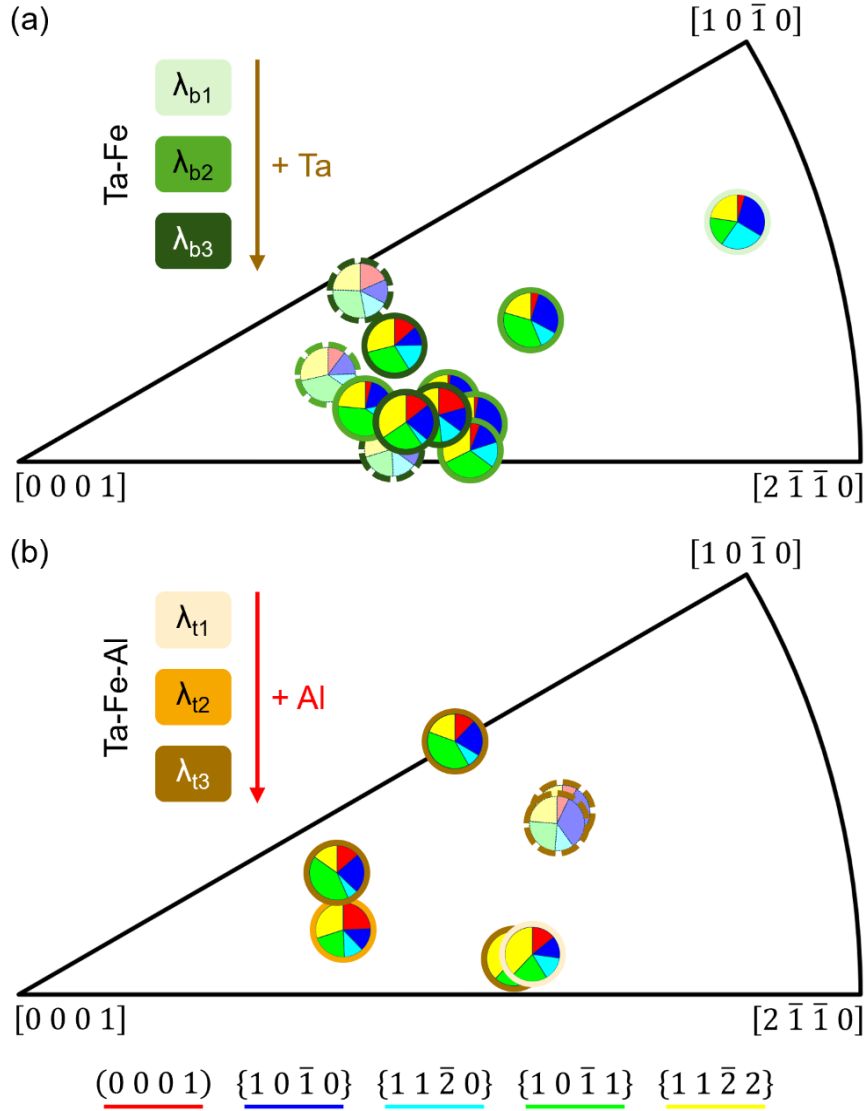

Figure 11: Relative activation frequency of the different activated slip planes for the (a) binary and (b) ternary Laves phase depending on the orientation marked within the IPF. The colour scheme for the different slip planes is given. The different compositions of the Laves phase samples are shown by different green and yellow shades for the binary and ternary Laves phase, respectively. With increasing darkening of the colours, the Ta content for the binary or the Al content for the ternary system increases. Orientations for which ten indents could be examined are displayed in bright colours with continuous frames, while orientation where only five to nine indents could be analysed are highlighted slightly transparent with a dashed frame.

For the binary and ternary Laves phase samples, mainly non-basal slip planes could be observed, see Figure 11. When comparing the observed relative activation frequency with the theoretical one that considers the number of equivalent planes for different slip systems (Table I), the favoured slip planes for the investigated Laves phase compositions can be determined. The possibility of double indexing individual slip lines and thereby influencing the statistics has of course to be included in this comparison. However, for the binary and ternary Laves phases samples with a total activation frequency of 119 % and 126 %, respectively, the different slip plane families are clearly distinguishable for a large number of the tested orientations (cf. supplementary material, section S2, Figure S2 and Figure S4). Therefore, it can be concluded that although pyramidal slip is most frequently observed, the CRSS is the lowest for basal slip followed by prismatic I and then the remaining slip systems. For the ternary system, the relative activation frequency of the prismatic I plane increases, which may indicate a lower yield stress due to the addition of Al or is simply attributable to the texture of the tested samples.

Our experimental observations agree well with the activated slip systems reported in literature for the hexagonal C14 Laves phase. Early studies on the C14 $MgZn_2$ prototype Laves phase already reported plastic deformation at elevated temperatures along the basal as well as non-basal planes, including the prismatic I, pyramidal I and pyramidal II plane [74–76]. These observations are also supported by more recent investigations of other C14 Laves phases by nanomechanical testing at room and elevated temperatures. Consistent with the findings in literature, no dominance of basal slip could be observed despite correspondingly favourable orientations [25, 26, 77, 78]. However, an increased occurrence of basal slip for θ close to 45° while it decreased for θ close to 0° and 90° was also pointed out [26, 77]. With a larger increase in Ta content, as for sample $λ_{b3}$, the proportion of basal slip is significantly higher for the analysed orientations with θ between 37° and 67° and further increases for an inclination of the basal plane of more than 45°, see Figure 3 (c), for which deformation along the basal plane then becomes the favoured mechanism. This is consistent with previous publications, finding that the presence of point defects, like anti-site atoms and vacancies for off-stoichiometric compositions, results in a decrease of the energy barrier of the synchroshear mechanism [79, 80], which is the most favourable mechanism for basal slip in Laves phases [36].

### 4.1.2 µ-phase

As for the Laves phases, the relative activation of the slip planes is also given for the binary and ternary µ-phases in the IPF, considering the full orientation definition by the Euler angles (Bunge), as illustrated in Figure 12.

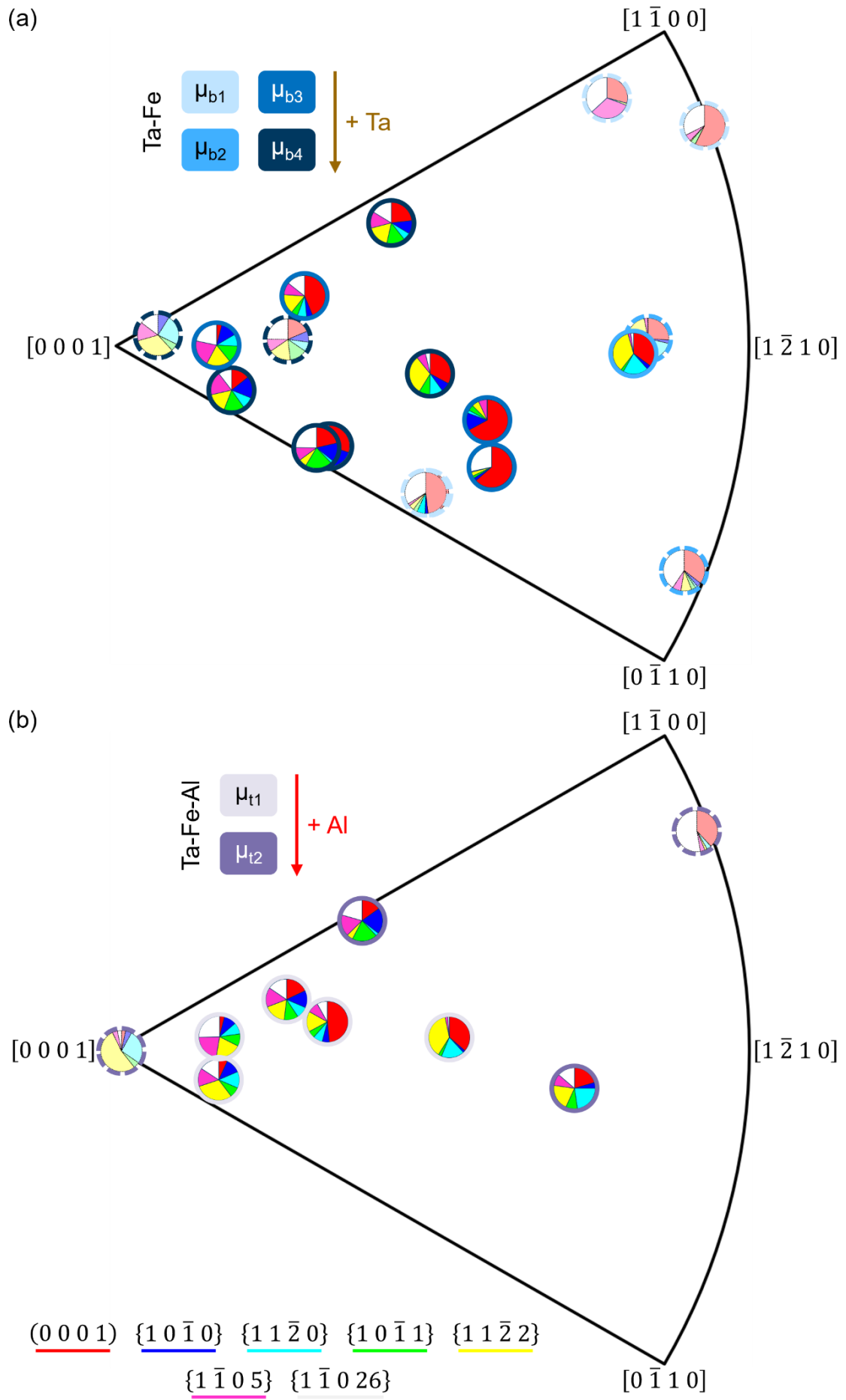

*Figure 12: Relative activation frequency of the different activated slip planes for the (a) binary and (b) ternary µ-phase depending on the orientation marked within the IPF. The colour scheme for the different slip planes is given. The different compositions of the µ-phase samples are shown by different blue and purple shades for the binary and ternary µ-phase, respectively. With increasing darkening of the colours, the Ta content for the binary or the Al content for the ternary system increases. Orientations for which ten indents could be examined are displayed in bright colours with continuous frames, while orientation where only five to nine indents could be analysed are highlighted slightly transparent with a dashed frame.*

Basal slip can be observed as the main deformation mechanism for the investigated µ-phase samples, as illustrated in Figure 12, which agrees with previous studies on plastic deformation in µ-phases. These were performed on different systems and composition ranges. For the Mo-Fe µ-phase [27] in the A-poor region and for the Nb-Co system [29] in the A-rich region, also mainly basal slip was observed. In addition, non-basal slip along the prismatic plane could be detected for the Mo-Fe system, while for the Nb-Co µ-phase, basal slip had to be suppressed in order to observe non-basal slip along the $\{1\,\bar{1}\,0\,5\}$ pyramidal plane [30] using TEM but to a much lesser extent than the basal slip lines clearly visible on the sample surface. Based on these findings and the TEM results of the investigated indents, besides the common hexagonal slip systems, also the $\{1\,\bar{1}\,0\,5\}$ plane and the $\{1\,\bar{1}\,0\,26\}$ A-B-A triple layer plane of the Ta-Fe(-Al) µ-phase were included in the slip trace analysis. When checking the relative activation frequency for the binary Ta-Fe µ-phase samples with different compositions it becomes apparent that slip along the added non-basal $\{1\,\bar{1}\,0\,5\}$ and $\{1\,\bar{1}\,0\,26\}$ planes is most frequently observed following basal slip. Indeed, these planes are observed in this careful analysis and can be clearly distinguished from the geometrically similar basal planes in selected orientations. However, with increasing last index of the Miller-Bravais indices, the inclination angle between the pyramidal and basal plane becomes smaller. Therefore, especially the $\{1\,\bar{1}\,0\,26\}$ A-B-A triple layer plane slip trace can appear similar as the one of the basal plane on the sample surface. Due to this similarity in the slip trace analysis double counting of possible slip lines happened more often for the µ-phase samples, see Figure S3 and Figure S5 (section S2 in the supplementary material). Furthermore, a higher activation frequency of basal slip for θ closer to 45° in the centre of the IPF could also be found in the µ-phase (especially for samples $µ_{b3}$, $µ_{b4}$ and $µ_{t1}$), which is explainable by the higher Schmid factor of the basal plane for these orientations and is consistent with our results on the Laves phases and the discussed literature.

The TEM results on the $Ta_7Fe_6$ µ-phase and therefore the same target composition as for sample $µ_{b3}$ agree well with the findings of the slip trace analysis. It was found that the double-indexed slip lines are predominantly the traces of the basal plane (see the many parallel slip bands below the indent in Figure 7 (b) and Figure 8), along which plastic deformation mainly occurs. Despite the selected grain orientation with θ equal to 37.7° for the lift-out of the TEM lamella favouring basal slip, slip bands along the $(0\,1\,\bar{1}\,5)$, $(0\,1\,\bar{1}\,26)$ and other pyramidal planes could also be observed (Figure 7 (b) and Figure 8). The activation of these non-basal planes might be explained by the complex stress field underneath an indent. Prismatic slip, on the other hand, as reported for the $Mo_6Fe_7$ µ-phase [27], was not found within the plastic zone of the analysed $Ta_7Fe_6$ µ-phase indent.

As for the Laves phase, alloying Al does only slightly influence the activated slip planes of the µ-phase towards non-basal slip. The TEM investigation of the µ-phase sample with 52 at.% Ta and an equal amount of Fe and Al with 24 at.% each, indicates slip along the basal plane as well as on the $(1\,\bar{1}\,0\,\bar{2})$ and $\{1\,\bar{1}\,0\,26\}$ planes, see Figure 9 (b). In this BF TEM montage image, the slip bands along the basal plane are mainly observed directly underneath the indent and to a greater number than slip bands along the $\{1\,\bar{1}\,0\,26\}$ plane. However, compared to the binary µ-phase sample, fewer slip bands identified as the basal plane were observed, while the amount of activated pyramidal planes increased significantly. Many $(1\,\bar{1}\,0\,\bar{2})$ slip bands are visible next to the right side of the indent and close to the sample surface. Planar defects on this plane were also mentioned in literature for a Ni-based µ-phase [81]. Further slip bands of planes that are not edge-on in Figure 9 (b) were analysed in Figure 10. The identified possible slip planes were not all included in the slip trace analysis, as they are partly very close and would therefore significantly increase the total activation frequency due to double counting. Furthermore, due to the crystal structure of the µ-phase some pyramidal planes only form families consisting of three and not six equivalent planes, whereby the energetically favourable ones were considered. Resulting from the complex stress field underneath an indent, the activation of the other (non-equivalent and less energetically favourable) slip planes is, however, also possible. Overall, in addition to basal slip, slip bands along various pyramidal planes were observed for the ternary µ-phase.

### 4.1.3  Effect of crystal structure and composition

Comparing the results of the TEM-supported slip trace analysis for the Laves and µ-phases, it is noticeable that despite the structural similarity of these two phases (with the Laves phase as a building block of the µ-phase), the activation of different slip planes can be observed. While non-basal slip predominates in the Laves phase, basal slip is much more pronounced in the µ-phase. The addition of Al seems to have no significant influence on the plasticity, only the proportion of non-basal slip slightly increases for both ternary TCP phases compared to the binary ones. However, changing the Ta content leads to more distinct changes for similar tested orientations. Within the binary Laves phase, increasing the Ta content causes an increase in basal slip. If the Ta content is increased within the homogeneity range of the µ-phase, the opposite can be observed. Here, the relative activation frequency of non-basal slip is highest for sample $µ_{b4}$, i.e. the sample with the highest Ta content, agreeing with the publication on the Nb-Co µ-phase that postulates that the synchroshear mechanism is not the main deformation mechanism along the basal plane for the µ-phase [28]. The effect of crystal structure and composition could only be identified in this kind of detailed analysis, in which the possible slip systems are determined for various (off-)stoichiometric compositions of the two TCP phases within the Ta-Fe(-Al) system.

## 4.2  Mechanical properties

Based on the data acquired in this work, we can compare stiffness and hardness across both crystal structures and with varying A atom concentrations (Ta content in the binary phases) and substitution of B atoms (Al content in the ternary phases), as summarised in Figure 13. Note that we do not resolve orientation dependence owing to the partly non-random orientation spread encountered. However, due to the three-dimensional stress state in indentation, orientation dependence is commonly small and only resolved in dedicated experiments [82], so that we anticipate a limited influence of differences in investigated textures but highlight where this may limit any interpretation.

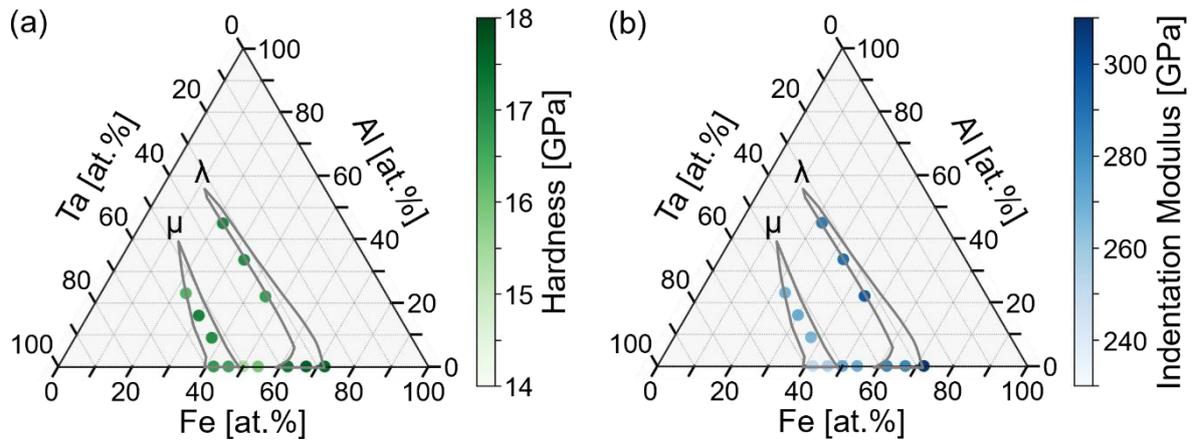

*Figure 13: Representation of the (a) hardness and (b) indentation modulus values (given in GPa) for all tested samples within a schematic of the ternary phase diagram. The phase range of the Laves (λ) and µ-phase according to the solidus projection is given, cf. [38, 39].*

### 4.2.1 Elasticity

#### 4.2.1.1 Crystal structure

The indentation modulus of the binary Ta-Fe system is clearly lower for the µ-phase than for the Laves phase see Figure 13 (b) and Figure 1. A similar observation was also made by Luo et al. [28] investigating the mechanical properties of the Nb-Co system and also observing a higher indentation modulus for the Laves phase than for any µ-phase composition. Since the µ-phase consists of a Laves phase building block alternately stacked with one of the $Zr_4Al_3$ type, the difference in the elastic behaviour might be explained by this in-between inserted $Zr_4Al_3$ structure. Studies in greater detail on this hypothesis have not been done so far, however, a previous work on the elastic properties of inorganic crystalline compounds [53] provides the elastic modulus of the prototype $Zr_4Al_3$ structure and a C14 Laves phase of the same system, the $ZrAl_2$ structure. The elastic or Young's modulus according to the Voigt-Reuss-Hill approach [60–62] is 189 GPa for the $Zr_4Al_3$ structure and 213 GPa for $ZrAl_2$ Laves phase. Although the interaction of the building blocks, the order and their bonding must also be considered for the mechanical properties of the µ-phase, these data provide a first indication that the indentation modulus of the µ-phase is lower than that of the Laves phase due to the lower stiffness of the inserted $Zr_4Al_3$ building block.

#### 4.2.1.2 Chemical composition

A decreasing trend of the indentation modulus can generally be observed with increasing Ta content in the binary system but also within the TCP phases itself for the different compositions, see Figure 13 (b) (cf. Figure 1). By adding more Ta, the smaller Fe atoms are replaced by the larger Ta atoms, especially within the triple layer as shown for the µ-phase [46, 47, 28]. Due to this replacement of the smaller by larger atoms, the interlayer and lattice spacings are expected to increase resulting in a reduced Peierls stress, a lower stiffness and therefore also a lower elastic modulus [28, 83, 84, 61]. Our DFT calculations suggest that the interlayer spacings $d_t$ and $d_{t-K}$ increase when adding more Ta from 0.369 and 1.637 Å, respectively, for $Ta_6Fe_7$ to 0.382 and 1.730 Å, respectively, for $Ta_7Fe_6$. Consistently, we observed a significant drop in the indentation modulus upon reaching the $Ta_7Fe_6$ composition (between samples $μ_{b2}$ and $μ_{b3}$), where the Fe atoms in the triple layer are fully replaced by Ta [46, 47, 28]. The decrease in modulus is again reproduced qualitatively by the DFT-calculated Young's modulus, as $Ta_6Fe_7$ and $Ta_7Fe_6$ have Young's moduli of 266 GPa and 241 GPa, respectively, leading to a decrease of 25 GPa or 10 %. While this drop is higher than the experimental value, the relatively less pristine coordination environment in the experimental samples, the contribution from sample textures, as well as the distinction between indentation modulus and Young's modulus, likely differentiate the experimental indentation modulus from the calculated Young's modulus.

Predicting the elastic modulus of ternary intermetallics is challenging, particularly where no isostructural binary compounds are available for first estimates assuming similar atomic volumes and bonding characteristics. Here, we alloyed Al into both the Laves and μ-phase, expecting it to replace Fe atoms on the B sites [82–84] and therefore resulting in an increased lattice spacing [85, 86]. Experimentally, we find that the elastic modulus does not show a strong dependence on the Al content within the ternary Fe-Ta-Al systems, neither in the Laves nor the μ-phase. Only a small reduction of the indentation modulus can be observed for the ternary Laves phase samples with a higher Al content. The Young's moduli range based on the three theoretically stable Ta-Fe-Al structural motifs, $Ta_4Fe_6Al_2$, $Ta_4Fe_3Al_5$ and $Ta_4Fe_2Al_6$, also shows that the stiffness changes only slightly with the Al content. Based on Hill's formulation, the range is only 12 GPa, which is relatively narrow. The stiffness of the investigated Ta-Fe TCP phases can therefore be estimated as constant upon alloying with Al as the third element. Whether this assumption will also hold true for other elements in the context of compositionally complex alloys containing these phases or whether phases with significant changes in local bonding and therefore stiffness occur, remains to be explored. Forming a greater understanding of the correlations between crystal structure, composition and modulus will aid both the understanding and prediction of precipitate formation in alloys as well as the design of reinforcement or bulk phases with desired stiffness and tuneable density.

### 4.2.2 Plasticity

#### 4.2.2.1 Laves phase

There are contradictory findings in literature on the influence of composition on the hardness of Laves phases. While some observe a hardening by off-stoichiometry [84, 87, 88], others report a softening [89–91]. In this work, the binary Ta-Fe Laves phase shows a constant hardness, although the average is slightly higher for the near-stoichiometric composition but well within the standard deviation. Softening is commonly explained in terms of the increased presence of point defects in form of vacancies and anti-site atoms on the A- and B-rich sides, respectively [18, 43–45]. These experimental observations of off-stoichiometric softening were also supported by recent publications on the C14 $CaMg_2$ [79] and C15 $CaAl_2$ [80] Laves phases, studying the influence of point defects on the energy barrier for synchro-Shockley dislocation mobility.

For the ternary Ta-Fe-Al Laves phase samples, where the Ta content was kept constant at 33 at.% near-stoichiometry, we found that also in terms of hardness, Al does not significantly alter properties. The replacement of Fe atoms by Al with a metallic radius between the one of Fe and Ta [40, 41, 92] therefore appears to have less influence on the hardness than changing the composition in the binary system towards off-stoichiometry.

*4.2.2.2  µ-phase*

In case of the binary µ-phase samples, the hardness does not follow a uniform trend, nor does it remain constant. Starting from the nearly stoichiometric $Ta_6Fe_7$ composition, a small drop of the average hardness (within standard deviation) towards increasing Ta content is observed. As the $Ta_7Fe_6$ composition is reached and exceeded, the hardness increases by approximately 1 GPa. Structurally, the softest sample, $µ_{b3}$, with 50 at.% Ta and Fe each lies in between the two stoichiometric $Ta_6Fe_7$ and $Ta_7Fe_6$ compositions in terms of composition, with presumably half of the Fe atoms in the triple layer replaced by Ta atoms [28]. Consequently, as with off-stoichiometric Laves phases, there are point defects in form of anti-site atoms leading to a hardness reduction [18, 89]. By further increase of the Ta content, the triple layer solely consists of the larger Ta atoms when reaching the stoichiometric $Ta_7Fe_6$ µ-phase with an increased hardness compared to the off-stoichiometric $µ_{b2}$ sample [28]. A previous study on the plasticity of different compositions of the Nb-rich $Nb_6Co_7$ µ-phase [29] by micropillar compression tests showed that the $A_7B_6$ ($Nb_7Co_6$) composition deforms more easily along the basal plane and is therefore softer than other compositions with a lower fraction of the A component. The stress field resulting from nanoindentation as performed in this work is, however, more complex than for uniaxial compression tests and when comparing the hardness of sample $µ_{b3}$ with sample $µ_{b4}$ with an even higher A atom content of 58 at.% Ta than the stoichiometric $A_7B_6$ ($Ta_7Fe_6$) composition, a further increase in hardness can be seen. Based on the observed intersection for the $Ta_7Fe_6$ µ-phase of basal and non-basal dislocations in Figure 8 (b) and (d) and in our previous work [30], it can be assumed that an increase in non-basal slip, as observed for sample $µ_{b4}$ (section 4.1.2), leads to a hindered dislocation movement due to an increased mutual blocking of the dislocations along different slip planes. And this blockage might be the explanation for the observed hardness change. In previous works on the plasticity of µ-phases, the compositional range, in which some of the B (Fe) atoms in the Kagomé net also have to be replaced by A (Ta) atoms and therefore affecting the packing of the unit cell, has so far not been studied in greater detail.

The nearly consistent hardness of the ternary µ-phase samples (Figure 1 and Figure 13 (a)) agrees with the findings on the ternary Laves phase samples. Studying the site occupation for ternary µ-phases is of course much more complex than for ternary Laves phases, however, Al is in general also expected to prefer the B sites [93, 94], which is why the nearly consistent hardness of the ternary µ-phases samples can also be explained by the partial atom replacement of Fe by Al without changing the site occupation preferences and while keeping the Ta content constant.

## 4.3　Scope for tailoring of Laves and µ-phase properties

A comparison of the mechanical properties of the binary and ternary Laves and µ-phases, respectively, shows that the addition of Al does not change the absolute values of hardness and stiffness significantly. This leads to the finding for Ta-Fe-Al that for the case in which one element is replaced by another one that normally prefers the same sites (in this case Fe and Al prefer the B sites), the complex crystal structure has a bigger influence on the mechanical properties of the TCP phase of the investigated system than the actual properties of the input elements themselves. Whether this holds true for a more general selection of elements remains to be explored, however, Al as a non-magnetic main group element is chemically distinct from Fe and the more common constituent atoms in complex alloys, which are largely transition metals. In principle, our results open the possibility of alloying with light elements, such as Al, to decrease the weight of the material without significantly changing the mechanical properties of the TCP phase.

For the Laves phase, we could observe that with increasing Ta content the frequency of basal slip increases without significant change in hardness. Within the µ-phase, on the other hand, the hardness and active slip systems can be affected significantly by means of the A:B compositional ratio. In particular, previous work across different systems has shown the following effects:

A:B ratio < 6:7: occurrence of prismatic and basal slip as dominant mechanisms (Mo-Fe [27]).

6:7 < A:B ratio < 7:6: occurrence of pyramidal and basal slip as dominant mechanisms with progressive softening on the basal plane with increasing A content (Nb-Co [29], Ta-Fe (this work)).

7:6 < A:B ratio: occurrence of pyramidal and basal slip as dominant mechanisms with increasing non-basal activation and concurrent hardening (Ta-Fe (this work)).

The observed effects may be considerable, with a reduction in CRSS by 75 % observed on the basal plane of the Nb-Co µ-phase across its homogeneity range, with the $Nb_7Co_6$ composition yielding at 1 GPa as opposed to 4 GPa measured for $Nb_6Co_7$ [29]. Future work will no doubt further explore the underlying mechanism changes and any induced difference in activation barrier. Most importantly, understanding the effect of varying A:B ratio more generally will also allow the prediction of active slip systems and approximate critical stresses in more complex alloys and precipitates that are not commonly available for individual testing, such as in superalloys and advanced steels.

Our present work contributes to this overall goal by systematically exploring the effects of crystal structure with a pure and partial Laves phase structure (µ-phase), as well as the variation of the A:B ratio and substitution on the B sites within a single ternary system. We observe a clear scope for tailoring density or corrosion resistance of these intermetallics at constant mechanical properties by substitution of Al as a light element on the B sites and manipulation of the active deformation mechanisms and resulting critical stresses by changing the A:B atom ratio in the µ-phase.

# 5  Conclusions

The large number of samples tested for different compositions and orientations within the Laves and µ-phase of the binary Ta-Fe and the ternary Ta-Fe-Al system, has significantly increased our understanding of the influence of these three factors on the mechanical properties and deformation mechanisms.

Our main conclusions are:

- In the binary Ta-Fe system, the crystal structure and chemical composition have a significant influence on the indentation modulus that is lower for the µ-phase than for the Laves phase and also decreases within the phases with increasing Ta content.

- A consistent, theoretical prediction of the Young's modulus between different phases using DFT is difficult due to the paramagnetic property of Fe, however, the decreasing trend in the binary µ-phase with increasing Ta content is consistent with the experimental results.

- The addition of Al to the analysed TCP phases does not significantly influence the mechanical properties at a constant A:B atom ratio.

- The hardness shows little variation across the crystal structures and compositions. A change in hardness between the $A_6B_7$ and $A_7B_6$ compositions remains to be explored in more detail in the context of a predicted reduced CRSS on the basal plane and increased activation of non-basal slip.

- The predominant slip planes differ between the Laves and µ-phase despite their structural similarity: plastic deformation mainly occurs via non-basal slip in the Laves and via basal slip in the µ-phase.

- Upon addition of Al in the ternary µ-phase, fewer slip bands are observed along the basal plane while pyramidal slip occurs more frequently.

- Furthermore, the $\{1\,\bar{1}\,0\,26\}$ plane, corresponding to the orientation of the A-B-A triplets of the triple layer was found as a new slip plane in the µ-phase in addition to the recently discovered $\{1\,\bar{1}\,0\,5\}$ plane.

- A comparison with two further binary systems containing both Laves and µ-phases reveals consistent trends in the properties and active deformation mechanisms, which provide a great scope for tailoring in particular density, corrosion resistance and hardness and manipulating the balance between basal and non-basal slip.

Due to the structural proximity of the considered phases with TCP phases of other binary and ternary systems, we hope to support the investigation and prediction of properties of these complex structures also in compositionally more diverse alloys and their TCP precipitates and to further enable the application of these phases as versatile structural or functional materials.

## 6　Declaration of competing interest

On behalf of all authors, the corresponding author states that there is no conflict of interest.

## 7　Data availability

Data will be made available on request.

## 8　Acknowledgements

The authors thank Hauke Springer (IEHK, RWTH Aachen University), Ingo Yawei Gao and Florian Alexander Busch (both IMM, RWTH Aachen University) for their help with the sample synthesis and preparation. This project has received funding from the European Research Council (ERC) under the European Union's Horizon 2020 research and innovation programme (grant agreement No. 852096 FunBlocks).

# Supplementary Material

## S1  Area function

To ensure relative comparability of the mechanical properties measured by nanoindentation despite progressive testing of hard materials and the resulting tip wear, the area function was calibrated on fused silica before each newly tested sample using the Oliver and Pharr approach [1, 2]. Therefore, an array of at least 16 indents was performed. The ideal area function is given by Equation (1):

$$A(h_c)_{ideal} = 24.5 * h_c^2 \tag{1}$$

with $h_c$ [nm] being the indentation contact depth. The non-ideal area function can also be calculated based on a defined number of area constants $C_x$ for curve-fitting with in this case five terms (x = 0, ..., 4) and again the indentation contact depth $h_c$, as given by Equation (2):

$$A(h_c) = C_0 * h_c^2 + C_1 * h_c + C_2 * h_c^{1/2} + C_3 * h_c^{1/4} + C_4 * h_c^{1/8} \tag{2}$$

The for calibration used area functions $A(h_c)$ are plotted over the indentation contact depth $h_c$ in Figure S14, in which also the ideal area function is displayed for comparison. It was found that the slope of the area function increases with increasing test number and therefore a higher amount of performed indentation tests for the used indentation tip as indicated by the black arrow in Figure S14. The area describing the tip geometry thus increases, indicating that the tip becomes rounded (tip wear).

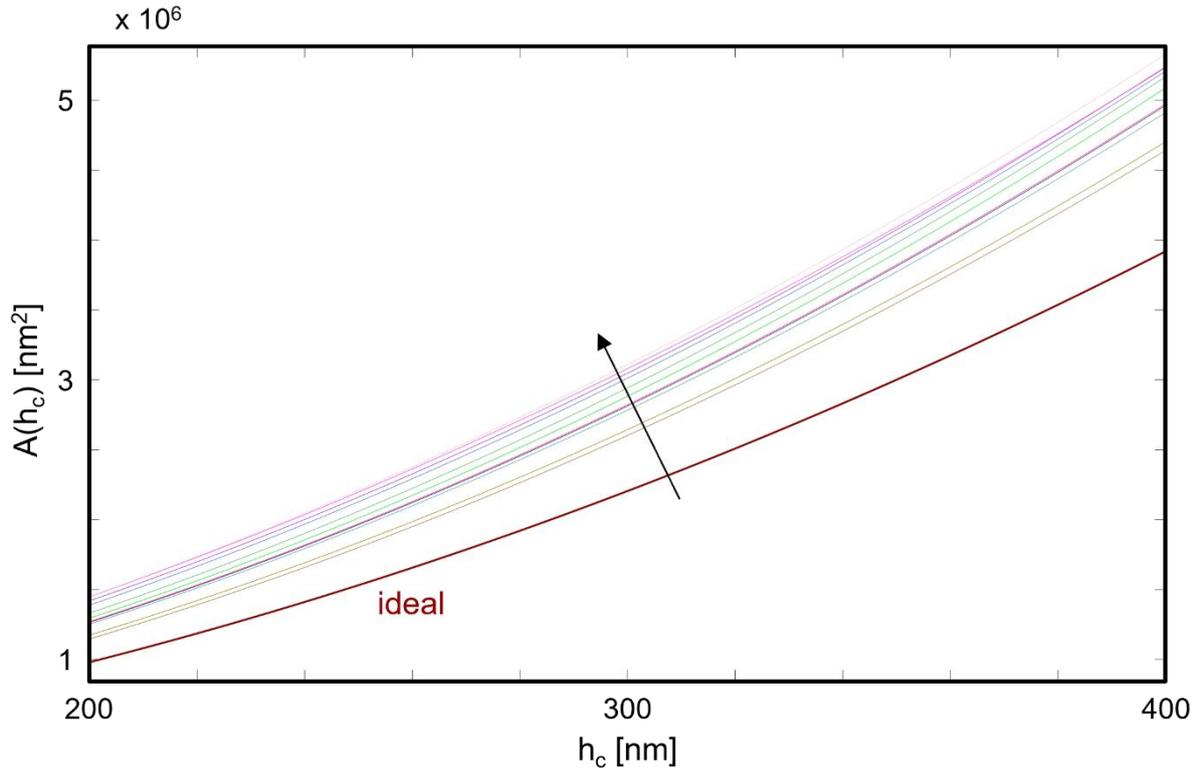

*Figure S14: Area functions $A(h_c)$ (given in nm$^2$) plotted over the indentation depth $h_c$ (given in nm). The ideal area function is highlighted in red and is the lowest curve in the diagram. The other coloured curves show an increasing trend in their slope with increasing test number (number of performed indents).*

## S2  Activation frequency versus second Euler angle θ

In addition to the representation of the relative slip plane activation within the IPF legend selected in section 3.2.2 in the manuscript, in the following the activation frequency of the analysed slip planes is plotted over the second Euler angle θ, which defines the inclination of the basal plane that is related to the tilting of the c-axis. It therefore functions as a good indicator for basal vs. non-basal slip. Based on this illustration, a direct comparison of the activated slip planes depending on the different compositions, phases and systems is possible. For the orientations where ten indents could be analysed, the corresponding columns are fully coloured and surrounded by a continuous line. Orientations for which fewer than ten indents could be analysed due to the grain size or the position of the indent array are also indicated with the frequencies of the activated slip planes in these diagrams in order to achieve the most complete dataset possible. However, they are shown transparently and have a dashed or dotted frame, depending on whether the data is available for five to nine or less than five indents, respectively. The colour scheme is given in each diagram. The five analysed slip planes for the Laves phase samples, the $(0\,0\,0\,1)$ basal, $\{1\,0\,\bar{1}\,0\}$ prismatic I and $\{1\,1\,\bar{2}\,0\}$ prismatic II as well as the $\{1\,0\,\bar{1}\,1\}$ pyramidal I and $\{1\,1\,\bar{2}\,2\}$ pyramidal II planes, are coloured in red, blue, turquoise, green and yellow, respectively. Additionally, the $\{1\,\bar{1}\,0\,5\}$ plane and the $\{1\,\bar{1}\,0\,26\}$ plane were included in the slip trace analysis for the μ-phase samples and are coloured in magenta and white, respectively.

The slip trace analysis results of the binary Ta-Fe Laves phase samples are shown in Figure S15. The activation frequency does not exceed 160 % and is on average 120 % for the three tested samples.

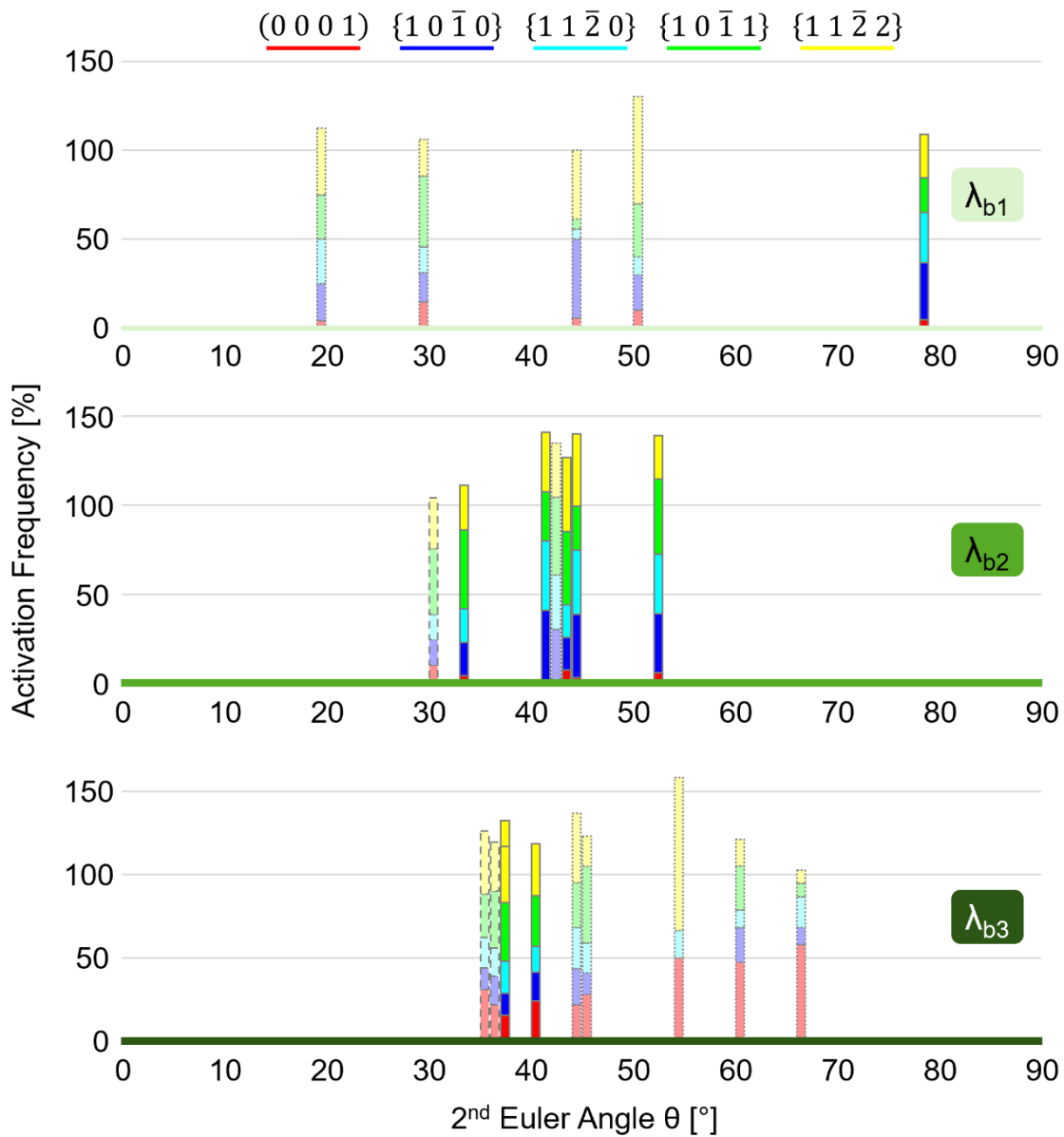

*Figure S15: Activation frequency of the analysed slip planes (given in %) plotted over the second Euler angle θ (given in °) for the binary Ta-Fe Laves phase samples $λ_{b1}$ with 28 at.% Ta, $λ_{b2}$ with 33 at.% Ta and $λ_{b3}$ with 38 at.% Ta (Fe balances all alloys to 100 at.%). The columns have different frames to illustrate the number of analysed indents per orientation: Continuous - ten indents, dashed – five to nine indents and dotted – less than five indents.*

In Figure S16 the activation frequency is plotted over θ for the four binary μ-phase compositions. The total activation frequency is higher than for the binary Laves phase and can reach values of up to almost 250 %, thus indicating that a larger number of solutions was considered for the actual slip lines.

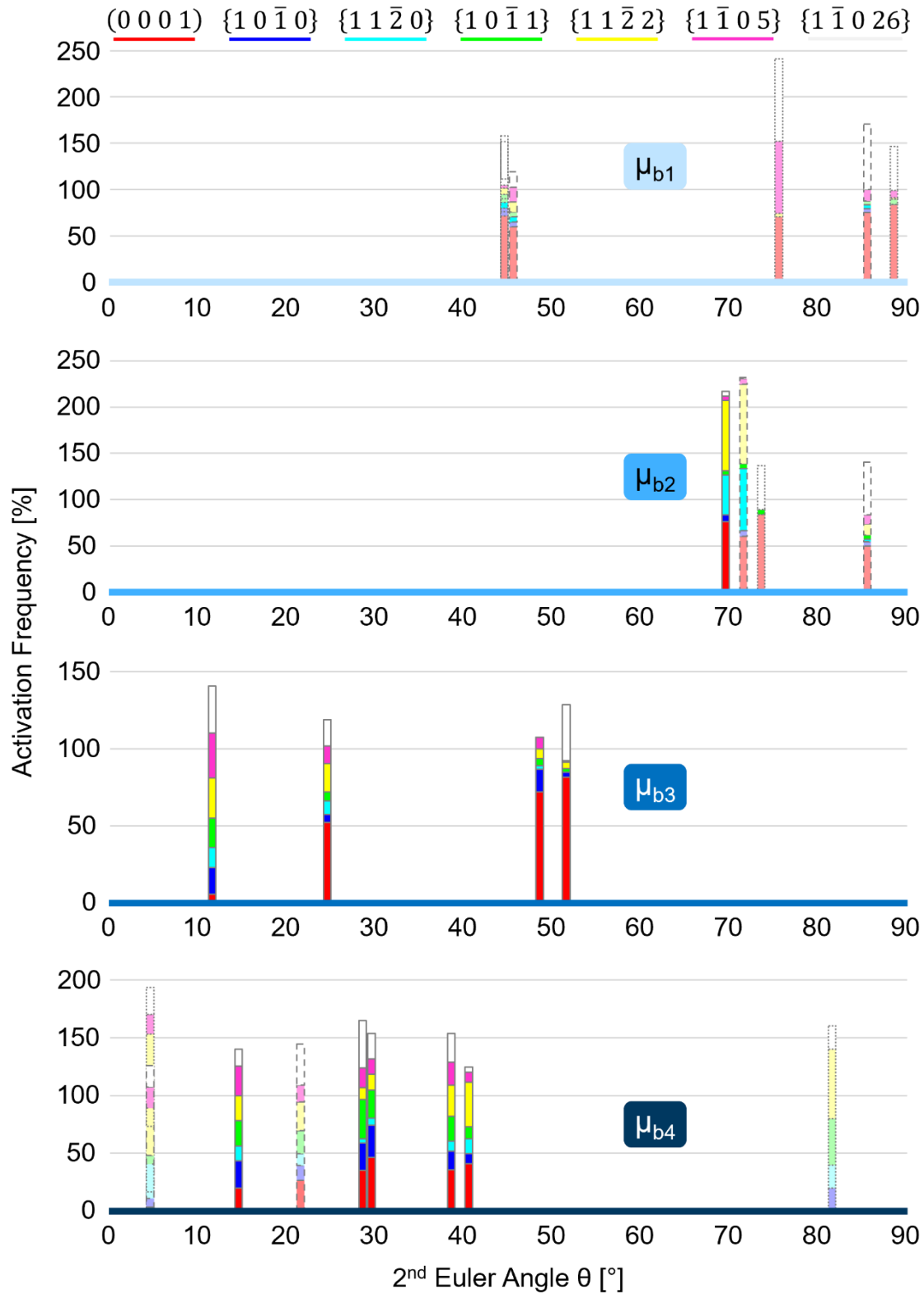

*Figure S16: Activation frequency of the analysed slip planes (given in %) plotted over the second Euler angle θ (given in °) for the binary Ta-Fe μ-phase samples μ$_{b1}$ with 46 at.% Ta, μ$_{b2}$ with 50 at.% Ta, μ$_{b3}$ with 54 at.% Ta and μ$_{b4}$ with 58 at.% Ta (Fe balances all alloys to 100 at.%). The legend for the colour scheme of the activated slip planes is given. The columns have different frames to illustrate the number of analysed indents per orientation: Continuous - ten indents, dashed – five to nine indents and dotted – less than five indents.*

For the ternary Laves phase sample, see Figure S17, the total activation frequency is again predominantly between 100 and 150 % (126 % on average).

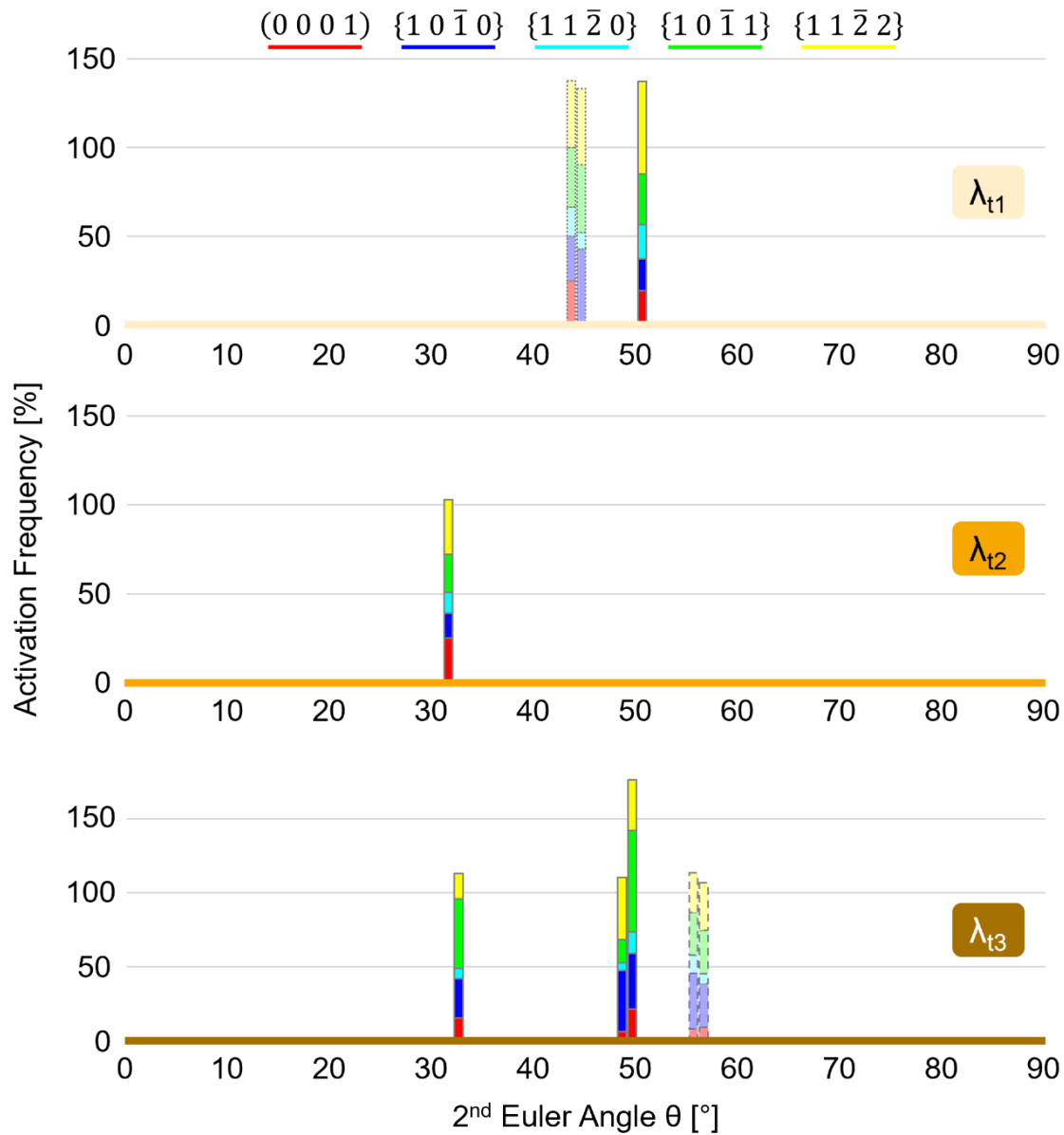

Figure S17: Activation frequency of the analysed slip planes (given in %) plotted over the second Euler angle θ (given in °) for the ternary Ta-Fe-Al Laves phase samples λ$_{t1}$ with 22 at.% Al, λ$_{t2}$ with 33.5 at.% Al and λ$_{t3}$ with 45 at.% Al (Ta content constant at 33 at.%, Fe balances all alloys to 100 at.%). The legend for the colour scheme of the activated slip planes is given. The columns have different frames to illustrate the number of analysed indents per orientation: Continuous - ten indents, dashed – five to nine indents and dotted – less than five indents.

The results of the slip trace analysis for the two ternary µ-phase samples are shown in Figure S18. As observed for the binary µ-phase, a high total activation frequency can again be observed for these samples, amounting to almost 300 % for some orientations of the µ$_{t2}$ sample. The proportion of basal slip does not exceed half of the total activation frequency in any orientation and is overall lower than in the binary µ-phase.

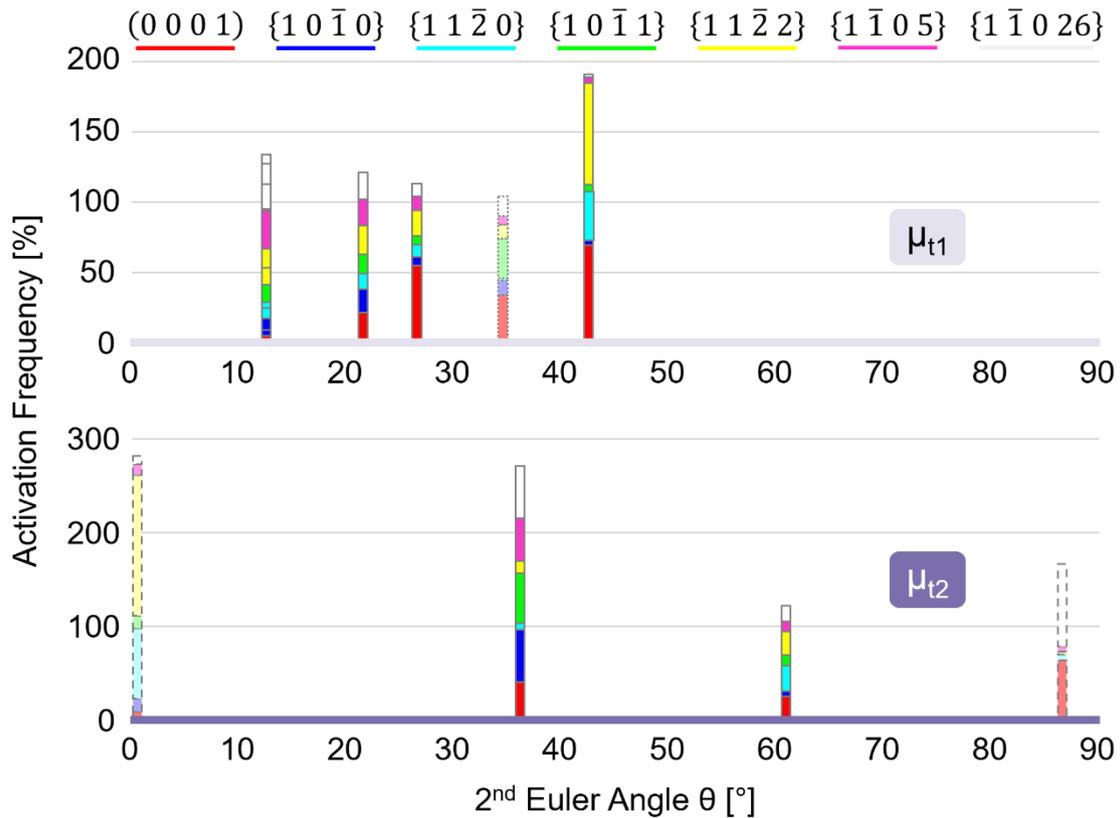

*Figure S18: Activation frequency of the analysed slip planes (given in %) plotted over the second Euler angle θ (given in °) for the ternary Ta-Fe-Al μ-phase samples μ$_{t1}$ with 9 at.% Al and μ$_{t2}$ with 16 at.% Al (Ta content constant at 54 at.%, Fe balances all alloys to 100 at.%). The legend for the colour scheme of the activated slip planes is given. The columns have different frames to illustrate the number of analysed indents per orientation: Continuous - ten indents, dashed – five to nine indents and dotted – less than five indents.*